%% file: sedentIII_revised.tex
\documentclass[twocolumn,usenatbib]{aa}
\usepackage{graphicx}
\usepackage{epsfig,graphicx,rotating,portland,longtable,natbib}
\usepackage{txfonts}
\usepackage{epstopdf}
\font\sevenrm=cmr7

\def\OII{[O~{\sevenrm II}]~}
\def\OIII{[O~{\sevenrm III}]~} 
\def\aro{{$\alpha_{\rm ro}$~}}

\def\ergj{{erg~cm$^{-2}$s$^{-1}$Jy$^{-1}$~}}

\def\fxfr{$f_{\rm x}/f_{\rm r}$~}

\begin{document}
   \title{The sedentary survey of extreme high-energy peaked BL Lacs \\ 
          III. Results from optical spectroscopy}

   \author{S. Piranomonte
          \inst{1}
          \and M. Perri\inst{2}
          \and P. Giommi\inst{2}
          \and H. Landt \inst{3}
          \and P. Padovani\inst{4}
          }
 
  \offprints{S. Piranomonte, \email{piranomonte@mporzio.astro.it}}

   \institute{INAF, Osservatorio Astronomico di Roma, 
              Via Frascati 33, 00040, Monte Porzio Catone (RM), Italy
	 \and 
	      ASI Science Data Center, c/o ESA-ESRIN, 
	      via Galileo Galilei, I-00044 Frascati, Italy
         \and
	      Harvard-Smithsonian Center for Astrophysics, 60 Garden Street, MS-29 Cambridge, MA 02138, USA
         \and 
	      European Southern Observatory, Karl-Schwarzschild-Str. 2, D-85748 Garching bei M\"unchen, Germany}

   \date{Received.. / Accepted... }

\abstract
{}
{The multi-frequency sedentary survey is a flux-limited, statistically well-defined 
sample of highly X-ray dominated (i.e., with a very high X-ray to radio flux ratio) BL Lacertae 
objects, which includes 150 sources. In this paper, the third of the series, we report the results 
of a dedicated optical spectroscopy campaign that, together with results from other independent 
optical follow-up programs, led to the spectroscopic identification of all sources in the sample.}
{We carried out a systematic spectroscopic campaign for the observation of all unidentified objects of the sample using the 
ESO 3.6m, the KPNO 4m, and the TNG optical telescopes.}
{We present new identifications and optical spectra for 76 sources, 50 of which are new BL Lac objects, 
18 are sources previously referred as BL Lacs but for which no redshift information was available, and 8 are 
broad emission-line AGNs. 
We find that the multi-frequency selection technique used to build the survey is highly 
efficient ($\sim 90\%$) in selecting BL Lacs objects.
We present positional and spectroscopic information for all confirmed BL Lac objects. Our data allowed us to 
determine 36 redshifts out of the 50 new BL Lacs and 5 new redshifts for the previously known objects. The redshift 
distribution of the complete sample is presented and compared with that of other BL Lacs samples. 
For 26 sources without recognizable absorption features, we calculated lower limits to the redshift using a method 
based on simulated optical spectra with different ratios between jet and galaxy emission.
For a subsample of 38 object with high-quality spectra, we find a correlation between the optical spectral slope, 
the 1.4 GHz radio luminosity, and the Ca H\&K break value, 
indicating that for powerful/beamed sources the optical light is dominated by the non-thermal emission from the jet.
}
{}

   
\authorrunning{Piranomonte et al.}
\titlerunning{The sedentary survey of extreme high-energy peaked BL Lacs III.}
\maketitle


\section{Introduction}

\label{intro}
BL Lacertae objects are strong radio-loud sources that constitute a rare subclass of active galactic 
nuclei (AGN) distinguished by peculiar and extreme properties, namely irregular and rapid variability, 
strong optical and radio polarization, lack of prominent emission lines, 
core-dominant radio morphology, and a broad continuum extending from the radio through high-energy $\gamma$-rays.
Their broad-band spectra are characterized, in a \mbox{$\nu\,f_\nu~\mathrm{vs}~\nu$} 
representation, by two emission peaks, the first located at IR/optical frequencies (but in 
several cases reaching the UV/X-ray band) and the second in the X-ray to $\gamma$-ray energy band.
The physical process that is believed to produce the low energy peak is synchrotron emission from 
relativistic electrons in the jet, while inverse Compton scattering by the same population of 
relativistic electrons is thought to be at the origin of the higher energy peak \citep[e.g.,][]{Ghi89}.
BL Lac objects are often divided into two classes according to the position of the synchrotron peak 
energy: low-energy peaked BL Lacs (LBLs), with the peak located at IR/optical wavelengths, and 
high-energy peaked BL Lacs (HBLs) with the synchrotron emission peaking in the UV/X-ray energy band 
\citep[e.g.,][]{Gio94, P95}.

A still open issue is that of the evolutionary properties of BL Lacs. It has in fact been reported (e.g. 
\citealt{Sti91,Mor91}) that BL Lacs have cosmological properties different 
from those of FSRQs and of all other type of AGNs. 
Although based on a few samples with rather small sizes, LBLs have been 
found to be consistent with being a non evolving population \citep{Sti91}, while HBLs seem 
to show a negative cosmological evolution; i.e. they are less numerous and/or less luminous at high 
redshift \citep{Mor91, Bade98, Rec00}.

Because of these extreme physical characteristics and of their unusual cosmological
evolution, BL Lacs have been the subject of intense research activity and 
observation campaigns from radio to TeV energies.  
HBLs are exceedingly rare with a density of less than one source (with 
radio flux larger than 3.5 mJy) every 100 square degrees \citep[e.g.,][]{Gio99}.
A classical approach, which requires identifying all sources in a radio-flux limited survey, 
would only reveal one HBL every $\sim$ 10,000 radio sources and would therefore be nearly 
impossible to pursue. All existing complete samples typically include less than 10 extreme HBLs, 
a subset far too small for any meaningful study.

The sedentary survey (\citealt*{Gio99}, hereafter referred to as Paper I) introduced a new 
multi-frequency, highly efficient approach to the discovery of HBLs. 
Using this method we have been able to assemble a well-defined, radio-flux-limited, 
sample including 150 objects (\citealt{paperII}, hereafter Paper II), which is 
currently the largest existing complete sample of high energy peaked BL Lacs. 

In order to identify candidate BL Lacs in the sedentary survey, we carried out a large 
optical spectroscopic campaign during which we observed all the unidentified sources 
of the sample, therefore making the survey 100\% spectroscopically identified.
In this paper, we present new optical spectroscopic data for 76 objects 
obtained during several observing runs at the KPNO 4m, the ESO 3.6 m and the TNG 3.6 m 
telescopes. 

The structure of the paper is as follows: in Section 2 we briefly describe the sedentary survey and its 
selection technique,  Section 3 discusses the results of our optical spectroscopy, Section 4 reviews some 
of the sample properties, in particular the redshift distribution. Our conclusions are summarized in Section 5.

Throughout this paper we have assumed cosmological parameters $H_0 = 50$ km s$^{-1}$
Mpc$^{-1}$ $q_0 = 0$. Spectral indices have been defined as $f_\nu \propto \nu^{-\alpha}$.

\section{The sedentary survey}

The Sedentary multi-frequency survey was designed to select a large and statistically well-defined sample of HBLs exploiting the fact that no other known source 
type has been found to possess such extreme spectral energy distribution. 
By imposing radio, optical, and X-ray flux ratios that are only
consistent with the unique spectral energy distribution of HBL sources, it is then possible to 
statistically select large samples of these rare sources.  

In the following we briefly recall the main definition criteria used to select 
this sample. We refer the reader to Papers I and II for details. The sample was extracted from a large set of radio and X-ray emitting sources
obtained by cross-correlating the ``ROSAT All Sky Survey Bright Source Catalog'' (RASS-BSC) of soft (0.1-2 keV) 
X-ray sources (18,811 sources, \citealt{Vog99}) with the NRAO VLA Sky Survey (NVSS) catalog of radio sources at 
1.4 GHz (1,807,316 sources, \citealt{Con98}). Optical magnitudes of the sources have been obtained from Palomar 
and UK Schmidt surveys through the APM and COSMOS services \citep{Irw94,Yen92}.

The following conditions have been imposed in order to ensure that the sample is complete 
above a radio flux limit of $f_{\rm r} \ge 3.5$ mJy 

\begin{enumerate}
\item $|b|>20^{\circ}$;
\item \fxfr $\ge 3\times 10^{-10}$ \ergj;
\item $\alpha_{\rm ro} > 0.2$;
\item $f_{\rm r} \ge 3.5$ mJy;
\item RASSBSC count rate $\ge 0.1$ cts/s;
\item $V \le 21$;
\end{enumerate}
\noindent
where $\alpha_{\rm ro}$ is the usual broad-band spectral index between
the radio (5GHz) and optical (5000 $\AA$) fluxes, and $V$ the visual 
apparent magnitude of the optical counterpart.

Condition 1) limits the survey area to high Galactic latitude regions where 
absorption due to Galactic $N_H$ is low; 
condition 2) imposes a very high \fxfr flux ratio
that, among radio loud sources, can be only reached by HBLs; condition 
3) removes radio quiet sources  from the sample, such as nearby Seyfert 
galaxies where the ratio between the unrelated radio and the X-ray flux accidentally 
satisfies condition 2);
conditions 4) and 5) and 6) ensure statistical completeness above 
$f_{\rm r} \ge 3.5$ mJy.

The sample has also been updated to include a few new BL Lacs that happened 
to be just below the \aro threshold of 0.2 used to separate BL Lacs from 
emission-line Seyfert galaxies \citep{Gio99}. In addition, a few spurious sources were found (consistent with the 
$\sim 15\%$ expected contamination from non BL Lac objects, \citealt{Gio99}) and removed from the 
sample.

Although the radio flux-limited sample was spectroscopically identified only at the $\sim40\%$ level when the first results 
were published (1999), its content was expected to include a high fraction ($\sim 85\%-90\%$) of HBL. This assumption is 
now confirmed both by the results of massive identification campaigns of X-ray sources discovered in the RASSBSC 
\citep{Bauer00, Schwope00, beck03} and by our spectroscopic identification of the remaining unclassified objects.

The sedentary survey sample is now completely identified and includes 150 HBLs (see Paper II). 
The full catalog is presented in Paper II and it is also available on-line at {\it http://www.asdc.asi.it/sedentary/}
\noindent where the broad-band spectral energy  distributions and the optical finding charts are also provided.


\section{Optical identification}
\label{ident}

The first identifications in the sample were obtained simply cross-correlating 
the precise NVSS positions with catalogs of known objects of different types. This first 
approach, discussed in Paper I, led to identification of 58 BL Lacs out of the 
original 155 HBL candidates.

In 1999 we started a systematic spectroscopic identification campaign 
to observe all the remaining HBL candidates or to obtain good-quality optical spectra 
of those objects already identified as BL Lacs but for which no redshift information was 
available. At the same time, \citet{Bauer00} and \citet{Schwope00} published the first results of the optical 
spectroscopic identification of bright X-ray sources in the ROSAT All Sky Survey, and 
\citet{beck00} reports optical identification of part of the Hamburg-RASS bright 
X-ray AGN sample (later published in \citealt{beck03}).
As expected, a significant fraction (about 40 objects) of the sedentary survey 
HBL candidates was found among these new classifications and thus 
the fraction of identified candidates in our sample significantly increased.
In August 2003 our spectroscopic program was completed with the identification 
of all candidates leading to a final sample of 150 spectroscopically identified HBLs.

\subsection{Spectroscopic observations}
\label{candidates}
\label{spec_obs}

\begin{figure*}
\vbox{
\psfig{figure=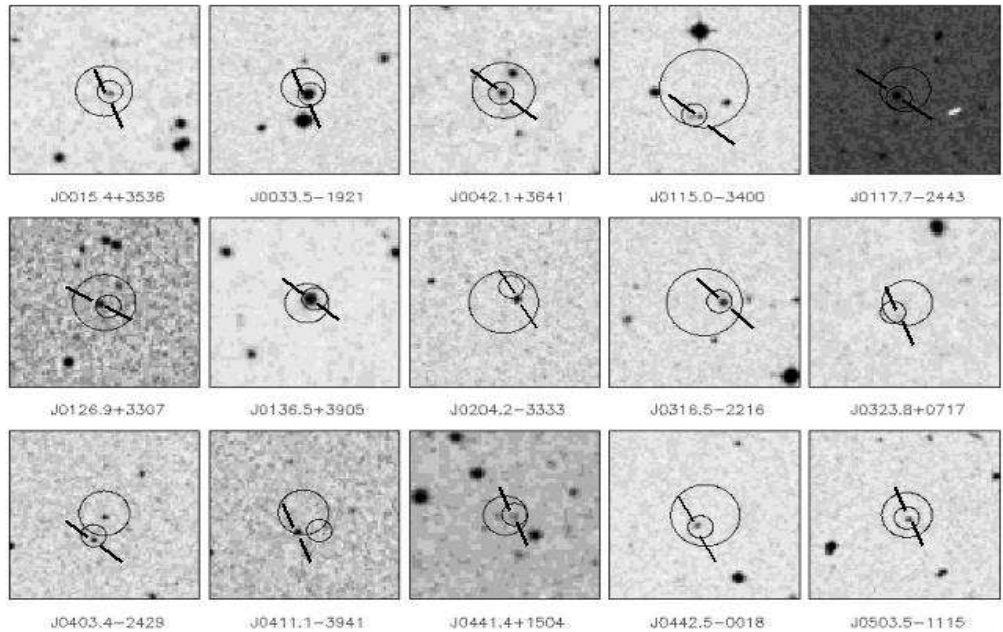,height=15cm,width=10cm,angle=-90}

}
\caption{A sample of optical finding charts of HBL sources observed by us. The X-ray and radio error circles 
(large and small, respectively) are shown.}
\label{finders_1}
\end{figure*}

%
%

Spectroscopic observations of the Sedentary sources still unidentified and of BL Lacs without redshift were carried 
out during the period 1999--2003 at the KPNO 4 m, at the ESO 3.6m and at the TNG telescopes.
The properties of grisms used in these runs are summarized in Table \ref{tab_instr}.

Finders of all the objects observed were taken from the on-line Digitized Sky Survey (DSS) in which both the 
X-ray and radio error circles have been plotted. This procedure is illustrated in Figs. \ref{finders_1} where 
it is shown that the accurate NVSS coordinate ($\la 3^{\prime\prime}$) in most cases permits a firm 
identification of the optical counterpart to be obtained.

\setcounter{table}{0}
\begin{table} [h]
\caption{Grism properties}
\label{tab_instr}
\begin{tabular}{llll} 
\hline 
\multicolumn{1}{c}{Telescope} & Grism & Dispersion & Range \\
& & (\AA~/pix) & (\AA~) \\
\hline
TNG 3.6 m & LR-B Grism 1& 2.8 & 3800$-$8000 \\ 
ESO 3.6 m & Grism 1 & 6.66 & 3185$-$10940 \\
& Grism 6 & 2.06 & 3860$-$8070 \\
& Grism 13 & 2.77 & 3685$-$9315 \\
KPNO 4 m & Grism 810 & 9.1 & 4300$-$10000 \\
\hline
\end{tabular}
\end{table}

Follow-up observations were made of 76 objects, including the unidentified BL Lacs candidates (58) and objects already 
classified as BL Lac from literature but without a redshift determination (18). In total we discovered 50 new BL Lac objects, 
according to the classification criteria employed in the classification scheme of \citet{Marcha96}; additionally, we determined 
5 new redshifts out of 18 already known BL Lacs. The remaining 8 objects observed by us were broad emission-line AGN that were 
excluded from the sample (see Table 3 of PaperII).

Adding the identifications from the literature published in \citet{paperII}, our technique is $\sim90\% $ efficient at selecting BL Lacs of the HBL type. 
In Table \ref{logs} we list the journal of observations for the 50 new BL Lacs discovered, for the 18 BL Lacs already known, and for 
the 8 broad emission lines AGNs. Columns are as follows: (1) name of the source, (2) telescope name, (3) date of observation, (4) exposure time (sec).

Standard data reduction was performed using different packages in IRAF\footnote{IRAF is distributed by the National Optical Astronomy Observatories, which are operated by the Association of the Universities for Research in Astronomy, Inc., under cooperative agreement with the National Science Foundation} to obtain 1-dimensional wavelength-calibrated extracted spectra. The data were bias-subtracted and flat-fielded using programs in the IRAF package {\it noao.imred.ccdred}, and the spectra were extracted, wavelength-, and flux-calibrated using programs in the package {\it noao.twodspec}.  A dereddening correction was applied to the data using the IRAF routine {\it noao.onedspec.deredden} and assuming Galactic values of extinction derived from 21-cm measurements \citep{stark92}. The spectra were wavelength-calibrated using an He-Ar (ESO), He (TNG), or He-Ne-Ar (KPNO) reference spectrum. The photometric standard stars used for the relative flux calibration are: Hiltner600 (TNG 02/2002), HR1544 (TNG 02/2002, TNG 09/2002), LTT3864 (ESO 05/2003), BD3326 (TNG 02/2003, KPNO 02/1999), CD3299 (ESO 07/2001), and LTT7379 (ESO 08/2003).

In general, we took two exposures for each object (except in few cases for which we have only one spectrum) in order to combine them and improve the signal-to-noise ratio (S/N) if necessary; this procedure allowed us to remove bad pixels and spurious features and to check the reliability of detected absorption and/or emission lines. In some cases dead pixels and cosmic rays were removed manually.

Most of the spectra were taken at parallactic angle, except in those cases where the radio/X-ray error circle contained two candidates, so a rotation of the slit was necessary. In those cases (8 objects) where we found two objects in the NVSS error box, we took the spectra of both objects. We always found that one object was a BL Lac and the other a star. 

\setcounter{table}{1}
\begin{table*} 
\begin{minipage}{150mm}
\caption{Log of observations}
\label{logs}
\begin{tabular}{lllrlllllr}
\hline 
Name & Observatory & Date & \multicolumn{1}{c}{Exp} & && Name & Observatory & Date & \multicolumn{1}{c}{Exp}\\
& & & \multicolumn{1}{c}{[sec]} & &&& & & \multicolumn{1}{c}{[sec]}\\ 
\hline 
\hline 

SHBL J001527.9$+$353639 & TNG 3.6 m  & 2002 Sep & 2400 &&& SHBL J125015.5$+$315559 & TNG 3.6 m  & 2002 Feb & 3600 \\ 
SHBL J003334.2$-$192133 & ESO 3.6 m  & 2001 July & 900  &&& SHBL J125341.1$-$393159 & ESO 3.6 m  & 2001 July & 2400 \\
SHBL J004208.0$+$364112 & TNG 3.6 m  & 2002 Sep & 2400 &&& SHBL J140630.1$-$393509 & ESO 3.6 m  & 2003 May & 3600 \\
SHBL J011501.9$-$340028 & ESO 3.6 m  & 2003 Aug & 3600 &&& SHBL J140630.2$+$123620 & ESO 3.6 m  & 2003 May & 4800 \\
SHBL J011747.0$-$244333 & ESO 3.6 m  & 2001 July & 2700 &&& SHBL J140659.2$+$164207 & ESO 3.6 m  & 2003 May & 1200 \\
SHBL J012657.2$+$330730 & TNG 3.6 m  & 2002 Sep & 2400 &&& SHBL J142739.5$-$252102 & ESO 3.6 m  & 2001 July & 2700 \\
SHBL J013632.5$+$390559 & TNG 3.6 m  & 2002 Sep & 600  &&& SHBL J143917.4$+$393243 & TNG 3.6 m  & 2003 Feb & 2700 \\
SHBL J020412.8$-$333342 & ESO 3.6 m  & 2001 July & 2400 &&& SHBL J144506.3$-$032612 & ESO 3.6 m  & 2003 May & 2400 \\
SHBL J031633.7$-$221611 & ESO 3.6 m  & 2001 July & 3000 &&& SHBL J150340.6$-$154113 & ESO 3.6 m  & 2001 July & 1800\\ 
SHBL J032350.7$+$071736 & ESO 3.6 m  & 2003 Aug & 3600 &&& SHBL J150637.0$-$054004 & ESO 3.6 m  & 2003 May & 2400\\   
SHBL J040324.5$-$242950 & ESO 3.6 m  & 2003 Aug & 1800 &&& SHBL J151041.0$+$333504 & KPNO 4 m   & 1999 Feb & 600 \\ 
SHBL J041112.2$-$394143 & ESO 3.6 m  & 2001 July & 3600 &&& SHBL J151618.6$-$152343 & ESO 3.6 m  & 2003 May & 1200 \\
SHBL J044127.4$+$150456 & ESO 3.6 m  & 2003 Aug & 2400 &&& SHBL J153311.3$+$185428 & TNG 3.6 m  & 2002 Sep & 2400 \\\  
SHBL J044230.1$-$001830 & TNG 3.6 m  & 2002 Sep & 1800 &&& SHBL J161204.6$-$043815 & ESO 3.6 m  & 2000 Aug & 2700 \\  
SHBL J050335.3$-$111507 & TNG 3.6 m  & 2003 Feb & 1800 &&& SHBL J161632.9$+$375602 & TNG 3.6 m  & 2002 Feb & 1800 \\ 
SHBL J050939.8$-$251403 & ESO 3.6 m  & 2003 Aug & 2400 &&& SHBL J163658.4$-$124837 & ESO 3.6 m  & 2001 July & 3600\\ 
SHBL J060714.3$-$251859 & ESO 3.6 m  & 2003 May & 2400 &&& SHBL J174702.3$+$493801 & TNG 3.6 m  & 2002 Sep & 3600 \\  
SHBL J062149.6$-$341149 & ESO 3.6 m  & 2003 May & 3600 &&& SHBL J175615.9$+$552217 & TNG 3.6 m  & 2002 Sep & 2400\\ 
SHBL J075124.9$+$173051 & KPNO 4 m   & 1999 Feb & 600  &&& SHBL J184822.3$+$653657 & TNG 3.6 m  & 2002 Sep & 1200\\ 
SHBL J075324.6$+$292132 & TNG 3.6 m  & 2002 Feb & 2400 &&& SHBL J203844.8$-$263633 & ESO 3.6 m  & 2001 July & 2400 \\
SHBL J092401.1$+$053345 & TNG 3.6 m  & 2002 Feb & 2400 &&& SHBL J204735.8$-$290859 & ESO 3.6 m  & 2001 July & 2700\\ 
SHBL J094355.5$-$070951 & ESO 3.6 m  & 2003 May & 3600 &&& SHBL J204921.7$+$003926 & ESO 3.6 m  & 2000 Aug & 1800 \\
SHBL J095224.0$+$750213 & KPNO 4 m   & 1999 Feb & 1200 &&& SHBL J205242.7$+$081038 & TNG 3.6 m  & 2002 Sep & 1800 \\
SHBL J095805.9$-$031740 & TNG 3.6 m  & 2002 Feb & 3600 &&& SHBL J213135.4$-$091523 & ESO 3.6 m  & 2001 July & 1200 \\
SHBL J101015.9$-$311908 & ESO 3.6 m  & 2003 May & 1200 &&& SHBL J213151.3$-$251558 & ESO 3.6 m  & 2001 July & 1800\\ 
SHBL J102243.8$-$011302 & ESO 3.6 m  & 2003 May & 2400 &&& SHBL J213852.5$-$205348 & ESO 3.6 m  & 2000 Aug & 1800 \\
SHBL J104651.4$-$253545 & ESO 3.6 m  & 2003 May & 1200 &&& SHBL J222253.8$-$175321 & ESO 3.6 m  & 2003 May & 2400\\ 
SHBL J111939.5$-$304720 & ESO 3.6 m  & 2003 May & 3600 &&& SHBL J224910.7$-$130002 & ESO 3.6 m  & 2001 July & 2700\\
SHBL J113444.5$-$172902 & ESO 3.6 m  & 2003 May & 3600 &&& SHBL J225147.3$-$320614 & ESO 3.6 m  & 2001 July & 2700\\
SHBL J113755.4$-$171042 & ESO 3.6 m  & 2003 May & 3600 &&& SHBL J230436.8$+$370507 & TNG 3.6 m  & 2002 Sep & 2400 \\
SHBL J114535.1$-$034001 & TNG 3.6 m  & 2003 Feb & 3600 &&& SHBL J230722.0$-$120518 & ESO 3.6 m  & 2003 May & 2400 \\
SHBL J123417.1$-$385635 & ESO 3.6 m  & 2001 July & 2400 &&& SHBL J231028.0$-$371909 & ESO 3.6 m  & 2001 July & 2400\\
SHBL J123511.0$-$140322 & ESO 3.6 m  & 1999 Mar & 1800 &&& SHBL J234333.8$+$344004 & TNG 3.6 m  & 2002 Sep & 3600\\ 
SHBL J124149.3$-$145558 & ESO 3.6 m  & 2003 May & 1200 &&& SHBL J235023.2$-$243603 & ESO 3.6 m  & 2001 July & 1200\\
\hline
1RXS J023727.6$-$26302  & ESO 3.6 m  & 2001 July& 2400 &&& 1RXS J170817.7$-$03493  & ESO 3.6 m  & 1999 Mar & 1800 \\
1RXS J123802.1$+$36164  & TNG 3.6 m  & 2002 Feb & 2400 &&& 1RXS J182042.7$+$38171  & TNG 3.6 m  & 2002 Sep & 2400 \\
1RXS J130350.5$-$39503  & ESO 3.6 m  & 2001 Jul & 2400 &&& 1RXS J212516.7$-$25553  & ESO 3.6 m  & 2000 Aug & 1800 \\
1RXS J133950.5$+$15593  & KPNO 4 m   & 1999 Feb & 1200 &&& 1RXS J222944.5$-$27553  & ESO 3.6 m  & 2001 July& 2400 \\
\hline 
\end{tabular}
\end{minipage}
\end{table*}

\subsection{Positional information}
\label{posinfo}
In Table \ref{pos} we list all the positional information, including the ones from RASSBSC and NVSS surveys, 
for the 76 sources observed during our identification campaign.
These form, together with the previously known sources, the complete sample published in Paper II. Columns are as follows: 
(1) name of the source, (2) and (3) RASSBSC position, (4) and (5) NVSS position, (6) and (7) position of the optical 
counterpart, which is taken from the on-line services APM and COSMOS and confirmed by our observations. 

\setcounter{table}{2}
\begin{table*}
\caption{Positional Information}
\label{pos}
\begin{center}
\begin{tabular}{lllllll}
\hline
Name & \multicolumn{2}{c}{RASSBSC Position} & \multicolumn{2}{c}{NVSS Position} & \multicolumn{2}{c}{Optical Position} \\
& RA (J2000) & DEC (J2000) & RA (J2000) & DEC (J2000) & RA (J2000) & DEC (J2000)  \\
\multicolumn{1}{l}{(1)} & \multicolumn{1}{c}{(2)} & \multicolumn{1}{c}{(3)} & \multicolumn{1}{c}{(4)} & \multicolumn{1}{c}{(5)} & \multicolumn{1}{c}{(6)} & \multicolumn{1}{c}{(7)} \\
\hline 
\hline
SHBL J001527.9$+$353639 &00 15 28.3 &$+$35 36 41&00 15 28.0 &$+$35 36 40.6 &00 15 27.9 &$+$35 36 39.1\\
SHBL J003334.2$-$192133 &00 33 34.6 &$-$19 21 29&00 33 34.3 &$-$19 21 33.7 &00 33 34.2 &$-$19 21 33.3 \\
SHBL J004208.0$+$364112 &00 42 08.1 &$+$36 41 15&00 42 08.2 &$+$36 41 12.9 &00 42 08.0 &$+$36 41 12.9\\
SHBL J011501.6$-$340027 &01 15 01.3 &$-$34 00 08&01 15 01.8 &$-$34 00 26.4 &01 15 01.6 &$-$34 00 27.0\\ 
SHBL J011747.0$-$244333 &01 17 46.6 &$-$24 43 29&01 17 46.9 &$-$24 43 35.0 &01 17 47.0 &$-$24 43 33.6 \\
SHBL J012657.2$+$330730 &01 26 57.1 &$+$33 07 30&01 26 57.0 &$+$33 07 27.3 &01 26 57.2 &$+$33 07 30.5\\
SHBL J013632.5$+$390559 &01 36 32.8 &$+$39 05 56&01 36 32.4 &$+$39 05 59.2 &01 36 32.5 &$+$39 05 59.6 \\
SHBL J020412.8$-$333342 &02 04 13.6 &$-$33 33 45&02 04 13.2 &$-$33 33 33.9 &02 04 12.8 &$-$33 33 42.5 \\
SHBL J031633.7$-$221611 &03 16 34.6 &$-$22 16 12&03 16 33.9 &$-$22 16 11.5 &03 16 33.7 &$-$22 16 11.5 \\
SHBL J032350.7$+$071737 &03 23 50.4 &$+$07 17 46&03 23 51.0 &$+$07 17 39.2 &03 23 50.7 &$+$07 17 37.0\\ 
SHBL J040324.5$-$242950 &04 03 24.0 &$-$24 29 31&04 03 24.6 &$-$24 29 46.6 &04 03 24.5 &$-$24 29 50.2\\
SHBL J041112.2$-$394143 &04 11 12.1 &$-$39 41 30&04 11 11.2 &$-$39 41 42.7 &04 11 12.2 &$-$39 41 43.8 \\
SHBL J044127.4$+$150456 &04 41 27.8 &$+$15 04 54&04 41 27.4 &$+$15 04 54.8 &04 41 27.4 &$+$15 04 56.0\\ 
SHBL J044230.1$-$001830 &04 42 29.8 &$-$00 18 23&04 42 30.0 &$-$00 18 31.1 &04 42 30.1 &$-$00 18 30.7 \\
SHBL J050335.3$-$111507 &05 03 35.5 &$-$11 15 04&05 03 35.5 &$-$11 15 06.1 &05 03 35.3 &$-$11 15 07.8\\
SHBL J050939.8$-$251403 &05 09 40.0 &$-$25 13 56&05 09 40.4 &$-$25 14 02.2 &05 09 39.8 &$-$25 14 03.0\\
SHBL J060714.3$-$251859 &06 07 14.2 &$-$25 18 55&06 07 14.3 &$-$25 18 59.6 &06 07 14.3 &$-$25 18 59.0\\
SHBL J062149.6$-$341149 &06 21 49.9 &$-$34 11 40&06 21 49.6 &$-$34 11 54.2 &06 21 49.6 &$-$34 11 49.8 \\
SHBL J075124.9$+$173051 &07 51 24.3 &$+$17 30 43&07 51 25.1 &$+$17 30 50.6 &07 51 24.9 &$+$17 30 51.0 \\
SHBL J075324.6$+$292132 &07 53 22.3 &$+$29 21 54&07 53 24.3 &$+$29 21 30.8 &07 53 24.6 &$+$29 21 32.0\\
SHBL J092401.1$+$053345 &09 24 01.1 &$+$05 33 50&09 24 01.2 &$+$05 33 42.7 &09 24 01.0 &$+$05 33 45.1 \\
SHBL J094355.5$-$070951 &09 43 55.3 &$-$07 09 43&09 43 55.5 &$-$07 09 52.5 &09 43 55.5 &$-$07 09 51.2 \\
SHBL J095224.0$+$750213 &09 52 25.8 &$+$75 02 16&09 52 23.7 &$+$75 02 13.2 &09 52 24.0 &$+$75 02 13.2 \\
SHBL J095805.9$-$031740 &09 58 06.4 &$-$03 17 29&09 58 06.1 &$-$03 17 38.3 &09 58 05.9 &$-$03 17 40.1 \\
SHBL J101015.9$-$311908 &10 10 15.9 &$-$31 19 09&10 10 15.9 &$-$31 19 06.5 &10 10 15.9 &$-$31 19 08.6\\
SHBL J102243.8$-$011302 &10 22 44.2 &$-$01 12 57&10 22 43.8 &$-$01 13 01.8 &10 22 43.8 &$-$01 13 02.5 \\
SHBL J104651.4$-$253545 &10 46 51.9 &$-$25 35 46&10 46 51.4 &$-$25 35 47.4 &10 46 51.4 &$-$25 35 45.2 \\
SHBL J111939.5$-$304720 &11 19 41.0 &$-$30 46 52&11 19 39.5 &$-$30 47 23.4 &11 19 39.5 &$-$30 47 20.2 \\
SHBL J113444.5$-$172902 &11 34 43.5 &$-$17 28 53&11 34 44.5 &$-$17 29 04.0 &11 34 44.5 &$-$17 29 02.6 \\
SHBL J113755.4$-$171042 &11 37 55.3 &$-$17 10 34&11 37 55.2 &$-$17 10 30.9 &11 37 55.4 &$-$17 10 42.8 \\
SHBL J114535.1$-$034001 &11 45 35.8 &$-$03 39 47&11 45 35.1 &$-$03 39 59.2 &11 45 35.1 &$-$03 40 01.4 \\
SHBL J123417.1$-$385635 &12 34 16.9 &$-$38 56 37&12 34 17.4 &$-$38 56 37.8 &12 34 17.1 &$-$38 56 35.0 \\
SHBL J123511.0$-$140322 &12 35 11.1 &$-$14 03 32&12 35 10.7 &$-$14 03 23.6 &12 35 11.0 &$-$14 03 22.6 \\
SHBL J124149.3$-$145558 &12 41 49.8 &$-$14 55 58&12 41 49.3 &$-$14 55 59.0 &12 41 49.3 &$-$14 55 58.0\\
SHBL J125015.5$+$315559 &12 50 15.0 &$+$31 56 04&12 50 15.7 &$+$31 56 04.0 &12 50 15.5 &$+$31 55 59.0 \\
SHBL J125341.1$-$393159 &12 53 41.2 &$-$39 32 00&12 53 41.3 &$-$39 31 59.8 &12 53 41.1 &$-$39 31 59.5 \\
SHBL J140630.1$-$393509 &14 06 30.3 &$-$39 35 08&14 06 30.3 &$-$39 35 10.4 &14 06 30.1 &$-$39 35 09.3\\
SHBL J140630.2$+$123620 &14 06 30.0 &$+$12 36 32&14 06 30.0 &$+$12 36 21.1 &14 06 30.2 &$+$12 36 20.4 \\
SHBL J140659.2$+$164207 &14 06 59.1 &$+$16 42 04&14 06 59.2 &$+$16 42 09.3 &14 06 59.2 &$+$16 42 07.2 \\
SHBL J142739.5$-$252102 &14 27 40.6 &$-$25 21 06&14 27 39.0 &$-$25 20 54.4 &14 27 39.5 &$-$25 21 02.0\\
SHBL J143917.4$+$393243 &14 39 17.7 &$+$39 32 48&14 39 17.4 &$+$39 32 42.4 &14 39 17.4 &$+$39 32 43.8 \\
SHBL J144506.3$-$032612 &14 45 05.8 &$-$03 26 13&14 45 06.3 &$-$03 26 13.5 &14 45 06.3 &$-$03 26 12.4 \\
SHBL J150340.6$-$154113 &15 03 42.9 &$-$15 41 07&15 03 40.6 &$-$15 41 17.8 &15 03 40.6 &$-$15 41 13.9 \\
SHBL J150637.0$-$054004 &15 06 36.4 &$-$05 40 10&15 06 37.0 &$-$05 40 06.5 &15 06 37.0 &$-$05 40 04.7 \\
SHBL J151041.0$+$333504 &15 10 40.8 &$+$33 35 15&15 10 42.0 &$+$33 35 08.7 &15 10 41.0 &$+$33 35 04.6 \\
SHBL J151618.6$-$152343 &15 16 18.6 &$-$15 23 47&15 16 18.2 &$-$15 23 44.0 &15 16 18.6 &$-$15 23 43.0\\
SHBL J153311.3$+$185428 &15 33 11.7 &$+$18 54 27&15 33 11.5 &$+$18 54 28.8 &15 33 11.3 &$+$18 54 28.5 \\
SHBL J161204.6$-$043815 &16 12 04.3 &$-$04 38 16&16 12 04.9 &$-$04 38 19.2 &16 12 04.6 &$-$04 38 15.9 \\
SHBL J161632.9$+$375602 &16 16 33.3 &$+$37 55 58&16 16 32.9 &$+$37 55 58.0 &16 16 32.9 &$+$37 56 02.7 \\
SHBL J163658.4$-$124837 &16 36 58.7 &$-$12 48 38&16 36 58.4 &$-$12 48 36.6 &16 36 58.4 &$-$12 48 37.3 \\
SHBL J174702.3$+$493801 &17 47 02.0 &$+$49 38 03&17 47 02.5 &$+$49 38 03.1 &17 47 02.3 &$+$49 38 01.7 \\
SHBL J175615.9$+$552217 &17 56 15.4 &$+$55 22 17&17 56 15.9 &$+$55 22 17.7 &17 56 15.9 &$+$55 22 17.7 \\
SHBL J184822.3$+$653657 &18 48 22.6 &$+$65 37 00&18 48 22.7 &$+$65 37 02.1 &18 48 22.3 &$+$65 36 57.3 \\
\hline
\end{tabular}
\end{center}
\end{table*}

\setcounter{table}{2}
\begin{table*}[t]
\caption{Positional Information -- \it{continued}}
\begin{center}
\begin{tabular}{lllllll}
\hline
Name & \multicolumn{2}{c}{RASSBSC Position} & \multicolumn{2}{c}{ NVSS Position} & \multicolumn{2}{c}{Optical Position} \\
& RA (J2000)& DEC (J2000)& RA (J2000)& DEC (J2000)& RA (J2000)& DEC (J2000) \\
\multicolumn{1}{l}{(1)} & \multicolumn{1}{c}{(2)} & \multicolumn{1}{c}{(3)} & \multicolumn{1}{c}{(4)} & \multicolumn{1}{c}{(5)} & \multicolumn{1}{c}{(6)} & \multicolumn{1}{c}{(7)} \\
\hline
\hline 
SHBL J203844.8$-$263633 &20 38 45.0 &$-$26 36 25&20 38 45.1 &$-$26 36 31.7 &20 38 44.8 &$-$26 36 33.9 \\
SHBL J204735.8$-$290859 &20 47 37.0 &$-$29 09 01&20 47 36.1 &$-$29 08 57.3 &20 47 35.8 &$-$29 08 59.5 \\
SHBL J204921.7$+$003926 &20 49 21.6 &$-$00 39 29&20 49 21.9 &$-$00 39 32.9 &20 49 21.7 &$-$00 39 26.9 \\
SHBL J205242.7$+$081038 &20 52 42.5 &$+$08 10 39&20 52 42.9 &$+$08 10 37.4 &20 52 42.7 &$+$08 10 38.2\\
SHBL J213135.4$-$091523 &21 31 35.5 &$-$09 15 25&21 31 35.4 &$-$09 15 22.8 &21 31 35.4 &$-$09 15 23.5 \\
SHBL J213151.3$-$251558 &21 31 51.7 &$-$25 16 01&21 31 51.5 &$-$25 15 58.7 &21 31 51.3 &$-$25 15 58.8 \\
SHBL J213852.5$-$205348 &21 38 52.8 &$-$20 53 54&21 38 52.8 &$-$20 53 45.3 &21 38 52.5 &$-$20 53 48.6 \\
SHBL J222253.8$-$175321 &22 22 53.8 &$-$17 53 17&22 22 54.1 &$-$17 53 24.0 &22 22 53.8 &$-$17 53 21.1 \\
SHBL J224910.7$-$130002 &22 49 11.0 &$-$13 00 05&22 49 11.0 &$-$12 59 57.4 &22 49 10.7 &$-$13 00 02.8 \\
SHBL J225147.3$-$320614 &22 51 46.8 &$-$32 06 14&22 51 47.5 &$-$32 06 17.0 &22 51 47.3 &$-$32 06 14.5 \\
SHBL J230436.8$+$370507 &23 04 37.1 &$+$37 05 06&23 04 36.6 &$+$37 05 07.3 &23 04 36.8 &$+$37 05 07.0 \\
SHBL J230722.0$-$120518 &23 07 22.5 &$-$12 05 20&23 07 22.0 &$-$12 05 18.5 &23 07 22.0 &$-$12 05 18.1 \\
SHBL J231028.0$-$371909 &23 10 26.9 &$-$37 19 26&23 10 28.2 &$-$37 19 08.5 &23 10 28.0 &$-$37 19 09.0 \\
SHBL J234333.8$+$344004 &23 43 32.4 &$+$34 39 57&23 43 33.8 &$+$34 40 00.8 &23 43 33.8 &$+$34 40 04.4 \\
SHBL J235023.2$-$243603 &23 50 23.6 &$-$24 35 52&23 50 23.4 &$-$24 36 05.4 &23 50 23.2 &$-$24 36 03.6\\
\hline
1RXS J023727.6$-$26302  &02 37 27.6 &$-$26 30 20.0 & 02 37 27.0 & $-$26 30 27.0 & 02 37 27.9 & $-$26 30 25.0\\
1RXS J123802.1$+$36164  &12 38 02.5 &$+$36 16 39.0 & 12 38 02.1 & $+$36 16 43.0 & 12 38 02.4 & $+$36 16 39.0 \\
1RXS J130350.5$-$39503  &13 03 50.6 &$-$39 50 53.0 & 13 03 50.5 & $-$39 50 35.0 & 13 03 50.6 & $-$39 50 43.0 \\
1RXS J133950.5$+$15593  &13 39 50.3 &$+$15 59 26.0 & 13 39 50.5 & $+$15 59 32.0 & 13 39 50.2 & $+$15 59 30.0 \\
1RXS J170817.7$-$03493  &17 08 18.0 &$-$03 49 16.0 & 17 08 17.7 & $-$03 49 37.0 & 17 08 17.9 & $-$03 49 16.0\\
1RXS J182042.7$+$38171  &18 20 43.6 &$+$38 17 01.0 & 18 20 42.7 & $+$38 17 11.0 & 18 20 43.5 & $+$38 17 00 \\
1RXS J212516.7$-$25553  &21 25 16.3 &$-$25 55 26.0 & 21 25 16.7 & $-$25 55 30.0 & 21 25 16.1 & $-$25 55 30.0\\
1RXS J222944.5$-$27553  &22 29 45.2 &$-$27 55 36.0 & 22 29 44.5 & $-$27 55 38.0 & 22 29 45.2 & $-$27 55 36.0\\
\hline 
\end{tabular}
\end{center}
\end{table*}

\subsection{Optical spectra}
\label{spectra}

In Appendix A, we present the spectra of the optical counterparts of
all BL Lacs observed by us (68). In Appendix B, for completeness, we also present 
the spectra of the 8 AGNs with emission lines excluded from the sample. 
All spectra were smoothed with a Gaussian filter of 3 pixel width.

The complete list of the observed BL Lacs objects (the 18 BL Lacs already known are 
marked with {\it d}), together with their properties,
are given in Table \ref{prop}, where the columns are as follows: (1)
source name, (2) unabsorbed $0.1 - 2.4$ keV X-ray flux;
(3) NVSS radio flux at 6 cm; (4) $V$ magnitude estimated from $O$ and
$E$ magnitudes obtained from the APM for the northern hemisphere and
from the COSMOS $B_J$ magnitudes as given in Paper I; (5) redshift,
computed, whenever possible, by taking the mean of the consistent
values derived from the absorption features; and (6) optical spectral
slope between the rest frame frequencies of $\OII \lambda3727$ and
$\OIII \lambda5007$. The Ca H\&K break value $C$ is given in column (7) and
was measured in spectra $f_{\lambda}$ versus $\lambda$ following
\citep{Dre87} as $C = 1- f_{-}/f_{+}$, where $f_{-}$ and $f_{+}$ are
the fluxes in the rest frame wavelength regions $3750-3950$~\AA~ and
$4050-4250$~\AA, respectively. We have considered the Ca H\&K break to
have reached its minimum value of zero when $f_{-}\ge f_{+}$. Its
$1\sigma$ error was calculated based on the S/N blueward and redward
of this feature. 
Finally, in column (8) we give the average S/N of the
spectrum around 5500~\AA~ measured in several $\sim200$~\AA~ intervals
and in (9) the $2\sigma$ upper limits on observed emission-line equivalent
widths are shown. For the latter we have assumed a rectangular emission line of
$FWHM=2000$ km/s centered at 5500~\AA.

\setcounter{table}{3}
\begin{table*}
\caption {Objects properties}
\label{prop}
\vspace{0.5cm}
\begin{tabular}{lcllllllll}
\hline
\multicolumn{1}{l}{Name} & F$_{0.1-2.4 keV}$ & F$_{20cm}$ & \multicolumn{1}{c}{Vmag}& \multicolumn{1}{c}{$z~^{(a)}$} &\multicolumn{1}{c}{$\alpha_{oiii}^{oii}$$^{(b)}$} &\multicolumn{1}{c}{Ca H\&K break $^{(c)}$} & \multicolumn{1}{c}{S/N} & \multicolumn{1}{c}{EW $^{(e)}$} \\
& [${\rm erg/cm^{2}/s}$] & [mJy] & & & & & &   \\
\multicolumn{1}{l}{(1)} & (2) & \multicolumn{1}{c}{(3)} & (4) & (5) & (6) & (7) & (8) & (9)\\
\hline
\hline
SHBL J001527.9$+$353639           & 3.45E-12& 11.2& 18.2& / 	 &/        &/               &  13 & $<$ 5.9  \\         
SHBL J003334.2$-$192133 $^{(d)}$  & 1.49E-11& 18.9& 16.1& 0.610? &1.04     &0               &  40 & $<$ 2.6  \\
SHBL J004208.0$+$364112 $^{(d)}$  & 3.68E-12& 12  & 17.9& /	 &/        &/               &  15 & $<$ 5.2  \\
SHBL J011501.9$-$340027           & 2.20E-12& 6.4 & 20.1& 0.482  &1.87     &$0.14 \pm 0.08$ &  15 & $<$ 3.1 \\
SHBL J011747.0$-$244333           & 3.82E-12& 10.3& 19  & 0.279  &4.03     &$0.30 \pm 0.05$ &  30 & $<$ 2.7  \\
SHBL J012657.2$+$330730           & 4.75E-12& 7.1 & 17.5& /	 &/        &/               &  12 & $<$ 6.5  \\
SHBL J013632.5$+$390559 $^{(d)}$  & 2.33E-11& 60.6& 15.4& /	 &/        &/               &  12 & $<$ 6.4  \\
SHBL J020412.8$-$333342           & 2.54E-12& 6.4 & 18.6& 0.617  &1.65     &$0.04 \pm 0.05$ &  27 & $<$ 2.9 \\
SHBL J031633.7$-$221611           & 2.44E-12& 4.4 & 19  & 0.228  &4.63     &$0.38 \pm 0.08$ &  27 & $<$ 2.9 \\
SHBL J032350.7$+$071737           & 6.69E-12& 4.5 & 20.3& /      &/        &/               &  10 & $<$ 7.4  \\
SHBL J040324.5$-$242950           & 5.01e-12& 7.4 & 20.1& 0.357  &2.67     &$0.09 \pm 0.12$ &  25 & $<$ 3.5 \\
SHBL J041112.2$-$394143           & 4.10E-12& 5.3 & 18.8& /	 &/        &/               &  12 & $<$ 6.5  \\
SHBL J044127.4$+$150456           & 3.93E-11& 14  & 19.8& 0.109	 &4.62     &$0.44 \pm 0.28$ &   8 & $<$ 8.4  \\
SHBL J044230.1$-$001830           & 4.06E-12& 4.2 & 20  & 0.449  &1.51     &$0.08 \pm 0.06$ &  17 & $<$ 4.1 \\
SHBL J050335.3$-$111507           & 1.29E-11& 10.5& 17.9& /      &/        &/               &   9 & $<$ 8.1 \\   
SHBL J050939.8$-$251403           & 2.13e-12& 4.5 & 20  & 0.264  &1.95     &$0.07 \pm 0.06$ &  30 & $<$ 2.7 \\
SHBL J060714.3$-$251859           & 4.13E-12& 12.2& 18.9& 0.275  &2.70     &$0.13 \pm 0.04$ &  28 & $<$ 2.9 \\
SHBL J062149.6$-$341149           & 4.47E-12& 8.8 & 18.7& 0.529  &0.86     &$0.02 \pm 0.04$ &  31 & $<$ 2.5\\
SHBL J075124.9$+$173051           & 3.85E-12& 10.5& 17.4& 0.185  &?        &?               &  21 & $<$ 3.6\\
SHBL J075324.6$+$292132           & 2.78E-12& 4.3 & 18.7& 0.161  &5.71     &$0.44 \pm 0.14$ &  21 & $<$ 3.5\\
SHBL J092401.1$+$053345 $^{(d)}$  & 5.15E-12& 7.6 & 18.4& /	 &/        &/               &  12 & $<$ 6.5  \\
SHBL J094355.5$-$070951           & 2.88E-12& 7.3 & 19.9& 0.433  &2.90     &$0.19 \pm 0.06$ &  22 & $<$ 3.1 \\
SHBL J095224.0$+$750213           & 4.83E-12& 12.4& 17.2& 0.179  &?        &?               &  21 & $<$ 3.5 \\
SHBL J095805.9$-$031740           & 2.82E-12& 7.6 & 19.6& /	 &/        &/               &  10 & $<$ 7.6\\
SHBL J101015.9$-$311908           & 2.83E-11& 74.4& 17.3& 0.143  &1.61     &0               &  41 & $<$ 1.8 \\
SHBL J102243.8$-$011302 $^{(d)}$  & 1.31E-11& 36.3& 17.2& /	 &/        &/               &  15 & $<$ 5.0 \\
SHBL J104651.4$-$253545           & 4.53E-12& 14.3& 19.2& 0.254  &2.27     &$0.08 \pm 0.05$ &  35 & $<$ 2.3\\
SHBL J111939.5$-$304720           & 3.72E-12& 9.6 & 19.5& 0.412  &2.11     &$0.08 \pm 0.05$ &  21 & $<$ 3.2\\
SHBL J113444.5$-$172902           & 3.64E-12& 5.0 & 19.7& 0.571  &1.81     &$0.02 \pm 0.05$ &  30 & $<$ 2.5\\
SHBL J113755.4$-$171042           & 4.72E-12& 5.3 & 18.9& 0.600  &0.66     &0               &  27 & $<$ 2.9\\
SHBL J114535.1$-$034001           & 7.96E-12& 19.7& 18.4& 0.167  &3.73     &$0.28 \pm 0.12$ &  13 & $<$ 5.5\\
SHBL J123417.1$-$385635           & 6.10E-12& 7.0 & 18.0& 0.236  &4.34     &$0.31 \pm 0.08$ &  23 & $<$ 3.4\\
SHBL J123511.0$-$140322           & 2.63E-12& 4.2 & 19.8& 0.407  &1.66     &$0.04 \pm 0.05$ &  20 & $<$ 3.4 \\
SHBL J124149.3$-$145558 $^{(d)}$  & 1.81E-11& 17.3& 17.3& /	 &/        &/               &  18 & $<$ 4.4  \\
SHBL J125015.5$+$315559           & 1.88E-12& 5.7 & 19.5& /	 &/        &/               &   4 & $<$18.6 \\
SHBL J125341.1$-$393159           & 1.96E-11& 50.1& 18.3& 0.179  &2.84     &$0.16 \pm 0.09$ &  22 & $<$ 3.5\\
SHBL J140630.1$-$393509           & 2.56E-12& 6.3 & 20.6& /	 &/        &/               &  20 & $<$ 3.9 \\
SHBL J140630.2$+$123620           & 4.04E-12& 8.8 & 19.8& /	 &/        &/               &  26 & $<$ 3.1 \\ 
SHBL J140659.2$+$164207 $^{(d)}$  & 5.43E-12& 8.4 & 18  & /	 &/        &/               &  16 & $<$ 4.9 \\
SHBL J142739.5$-$252102           & 4.70E-12& 3.7 & 18.9& 0.318  &5.12     &$0.46 \pm 0.13$ &  32 & $<$ 2.7 \\
SHBL J143917.4$+$393243 $^{(d)}$  & 1.79E-11& 42.9& 16.6& 0.344  &1.03     &0               &  29 & $<$ 2.3\\
SHBL J144506.3$-$032612 $^{(d)}$  & 7.80E-12& 21.8& 17.4& /	 &/        &/               &  17 & $<$ 4.5 \\
SHBL J150340.6$-$154113 $^{(d)}$  & 2.39E-11& 5.9 & 17.5& /	 &/        &/               &  14 & $<$ 5.6 \\
SHBL J150637.0$-$054004           & 5.03E-12& 15.3& 19.5& 0.518  &1.88     &$0.05 \pm 0.06$ &  28 & $<$ 2.6\\
SHBL J151041.0$+$333504           & 4.45E-12& 9.1 & 17.0& 0.112  &?        &?               &  18 & $<$ 3.9\\
SHBL J151618.6$-$152343 $^{(d)}$  & 1.46E-11& 8.6 & 18.7& /	 &/        &/               &  20 & $<$ 4.0 \\
SHBL J153311.3$+$185428 $^{(d)}$  & 1.43E-11& 23.0& 17.7& 0.305  &1.41     &$0.015 \pm 0.040$& 22 & $<$ 2.8 \\
SHBL J161204.6$-$043815           & 4.40E-12& 4.3 & 18.9& /	 &/        &/               &  11 & $<$ 6.8 \\
SHBL J161632.9$+$375602           & 1.47E-12& 4.6 & 18.7& 0.204  &4.18     &$0.27 \pm 0.10$ &  16 & $<$ 4.8\\
SHBL J163658.4$-$124837           & 8.74E-12& 26.3& 20.1& 0.246  &3.54     &$0.20 \pm 0.03$ &  36 & $<$ 2.2\\
SHBL J174702.3$+$493801           & 2.81E-12& 7.9 & 19.9& 0.460  &2.26     &$0.13 \pm 0.07$ &  20 & $<$ 3.6\\
SHBL J175615.9$+$552217 $^{(d)}$  & 1.48E-11& 16.9& 17.6& /	 &/        &/               &  12 & $<$ 6.4 \\
SHBL J184822.3$+$653657           & 4.29E-12& 9.7 & 18.3& 0.364  &2.86     &$0.11 \pm 0.09$ &  18 & $<$ 3.7\\
\hline
\end{tabular}
\end{table*}

\setcounter{table}{3}
\begin{table*}
\caption {Objects Properties -- \it continued}
\vspace{0.5cm}
\begin{tabular}{lcllllllll}
\hline
\multicolumn{1}{l}{Name} & F$_{0.1-2.4 keV}$ & F$_{20cm}$ & \multicolumn{1}{c}{Vmag}& \multicolumn{1}{c}{$z~^{(a)}$} & \multicolumn{1}{c}{$\alpha_{oiii}^{oii}$$^{(b)}$} &\multicolumn{1}{c}{Ca H\&K break $^{(c)}$} & S/N & \multicolumn{1}{c}{EW$^{(e)}$} \\
& [${\rm erg/cm^{2}/s}$] & [mJy] & & & & & &\\
\multicolumn{1}{l}{(1)} & (2) & \multicolumn{1}{c}{(3)} & (4) & (5) & (6) & (7) & (8) & (9)\\
\hline
\hline
SHBL J203844.8$-$263633           & 4.73E-12& 5.7 & 18.5& 0.437  &2.49     &$0.15 \pm 0.07$ &  21 & $<$ 3.3 \\
SHBL J204735.8$-$290859           & 3.97E-12& 10.7& 19.3& 0.333  &2.28     &$0.09 \pm 0.02$ &  33 & $<$ 2.6\\
SHBL J204921.7$+$003926           & 4.85E-12& 6.0 & 18.1& 0.256  &3.25     &$0.21 \pm 0.14$ &  18 & $<$ 4.4\\
SHBL J205242.7$+$081038           & 4.90E-12& 6.2 & 19.6& /	 &/        &/               &  12 & $<$ 6.3 \\
SHBL J213135.4$-$091523 $^{(d)}$  & 1.58E-11& 43.6& 16.6& 0.449? &0.91     & 0               &  26 & $<$ 2.7\\
SHBL J213151.3$-$251558           & 6.04E-12& 11.0& 17.3& /	 &/        & /               &  18 & $<$ 4.2 \\
SHBL J213852.5$-$205348 $^{(d)}$  & 1.33E-11& 11.5& 17.9& 0.290  &1.22     & 0               &  21 & $<$ 3.9 \\
SHBL J222253.8$-$175321           & 2.88E-12& 5.7 & 19.4& 0.297  &2.10     & $0.06 \pm 0.05$ &  25 & $<$ 3.3\\
SHBL J224910.7$-$130002 $^{(d)}$  & 9.73E-12& 7.5 & 18.9& /	 &/        & /               &  13 & $<$ 5.9 \\
SHBL J225147.3$-$320614 $^{(d)}$  & 3.61E-12& 3.6 & 19.0& /	 &/        & /               &  19 & $<$ 4.0 \\
SHBL J230436.8$+$370507 $^{(d)}$  & 1.85E-11& 23.1& 17.8& /	 &/        & /               &  11 & $<$ 7.2 \\
SHBL J230722.0$-$120518           & 2.94E-12& 7.3 & 18.5& /	 &/        & /               &  14 & $<$ 5.4 \\
SHBL J231028.0$-$371909           & 1.97E-12& 6.3 & 17.8& /	 &/        & /               &  21 & $<$ 3.8 \\
SHBL J234333.8$+$344004           & 1.51E-11& 35  & 20.1& 0.366  &2.19     & $0.06 \pm 0.05$ &  31 & $<$ 2.2\\
SHBL J235023.2$-$243603           & 2.14E-12& 6.7 & 16.5& 0.193  &5.02     & $0.40 \pm 0.19$ &  24 & $<$ 3.1\\
\hline
\end{tabular}
\\

$^{(a)}$ uncertain redshifts are marked with a ``?''.\\
$^{(b)}$ Impossible measure of $\alpha_{oiii}^{oii}$ (the $\OII \lambda3727$ is located in noise) is marked with a ``?''.\\
$^{(c)}$ we consider the Ca H\&K break to have reached its minimum value of zero when the flux blue-ward of this feature is equal to or larger than the one red-ward. The Ca H\&K break located in noise are marked with a ``?''.\\
$^{(d)}$ sources already classified in literature as BL Lacs (see Paper II). \\
$^{(e)}$ 2$\sigma$ upper limit of observed EW, see text for details. \\
\end{table*}

\subsection{Redshift determination}
\label{redshift_new}

In the optical band, the spectrum of a BL Lac is made up of two main 
components: (i) the amplified non-thermal jet emission, which follows a
power-law of the form $f_{\nu} \propto \nu^{-\alpha}$, with $\alpha$
the spectral index, and (ii) thermal emission from the host galaxy,
normally a luminous elliptical \citep[e.g.,][]{Wur96, Urry00}. The
emission-line regions in BL Lacs are, by definition, only very weak
or absent (see Section 3.5. for more details), which means that their
redshift determination relies strongly on the detection of galaxy
absorption features. This, however, is only possible if the jet is
weak relative to the galaxy, i.e., only for low-luminosity BL Lacs. In
strongly beamed sources, the jet with its featureless spectrum will 
dilute any galaxy absorption features beyond recognition
\citep*{L02}. In the sample identified by us, the redshift was 
determined based on emission lines only for 8 objects (Appendix B). For most
sources (41 objects) we have used the absorption features typical of
ellipticals, the strongest of which are summarized in Table \ref{abs_ellipt}. In a
considerable fraction of our objects (36\% or 27/76 objects), however, we observe
only a featureless spectrum for which no reliable redshift
determination is possible. This fraction reduces to 23\% (39/169
objects), if we consider the entire sample of HBLs (150 sources, Paper II) 
and emission-line AGN (19 sources, see Paper II).

\setcounter{table}{4}
\begin{table} [h]
\caption{Absorption features typical of ellipticals}
\label{abs_ellipt}
\begin{tabular}{lc} 
\hline 
\multicolumn{1}{l}{Absorption Feature} & \multicolumn{1}{c}{$\lambda$ (\AA)} \\
\hline
FeII     &  2402        \\   
BL       &  2538        \\   
FeII     &  2609        \\   
MgII     &  2800        \\   
MgI      &  2852        \\   
FeI      &  3000        \\   
BL       &  3096        \\   
BL       &  3580        \\	   
Ca H\&K   &  3934\&3968 \\
G band   &  4300        \\   
H$\beta$ &  4862        \\   
MgIb     &  5174        \\   
NaD      &  5891        \\   
\hline
\end{tabular}
\end{table}

\begin{figure}[h]
\centering
\includegraphics[height=8cm, width=8cm]{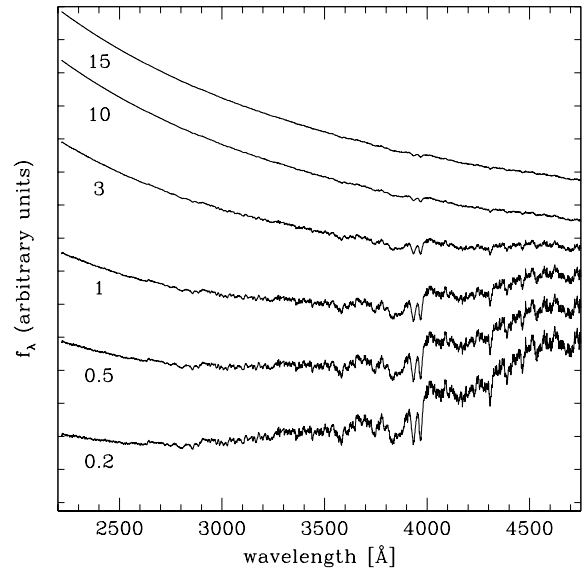}
\caption{Simulated BL Lac spectra $f_{\lambda}$ vs. $\lambda$ for a
jet of optical spectral slope $\alpha=1$ and of increasing flux
relative to the underlying host galaxy. The assumed jet/galaxy ratios
(defined at 5500 \AA) are from bottom to top: 0.2, 0.5, 1, 3, 10, 15.}
\label{highz}
\end{figure}

In order to still be able to conduct meaningful cosmological studies
with our sample, we have developed a method of determining 
lower limits on the redshift of sources without recognizable
absorption features. The simulations of \citet*{L02} of low-redshift
BL Lac spectra (see their Fig. 1) show that BL Lacs are expected to
become featureless for jet/galaxy ratios $\ga10$ (defined at 5500
\AA). However, the absorption features of ellipticals present at
larger rest-frame wavelengths (i.e., redward of the Ca H\&K break) are
considerably stronger than the ones found at smaller rest-frame
wavelengths (see Fig. \ref{highz}), which means that the situation
will be different for high-redshift BL Lacs. For these we expect the
spectrum to become featureless at even smaller jet/galaxy ratios.

To determine this value we extended the simulations of
\citet*{L02} to smaller rest-frame wavelengths. Our results are shown
in Fig. \ref{highz} representative of a jet of optical spectral
slope $\alpha=1$. As soon as the Ca H\&K break
moves out of the ``useful'' optical window (i.e., lies at observed
wavelengths $\ga6500 \AA$, corresponding to redshifts of $z\ga0.65$,
where prominent telluric absorption bands dominate), a redshift
determination based on absorption features is expected to be already 
impossible for sources with jet/galaxy ratios
$\ga1$. Therefore, a featureless, power-law like spectrum indicates
that either the BL Lac is at low redshift ($z\la0.65$) and 
strongly beamed (i.e., its jet/galaxy ratio is high) or it is at high
redshifts ($z\ga0.65$) where it can be both moderately or highly
beamed.

\setcounter{table}{5}
\begin{table*}
\caption {Featureless Sedentary HBL.}
\label{featureless}
\vspace{0.5cm}
\begin{tabular}{llcllllc}
\hline
Name &  Vmag & \multicolumn{1}{c}{z} & & & Name & Vmag & \multicolumn{1}{c}{z}\\
& & \multicolumn{1}{c}{[lower limit]} &&& & & \multicolumn{1}{c}{[lower limit]}  \\
(1) & (2) & \multicolumn{1}{c}{(3)} & & & (1) & (2) & \multicolumn{1}{c}{(3)}\\
\hline 
(our sample) & & &&& (literature)& & \\
\hline 

SHBLJ001527.9+353639            & 18.3  & 0.650 &&& SHBLJ005816.8+172310$^{(b)}$    &  19.2 & /      \\ 
SHBLJ004208.0+364112            & 18.0  & 0.623 &&& SHBLJ014040.7$-$075848          &  18.3 & 0.650  \\
SHBLJ012657.2+330730            & 17.5  & 0.497 &&& SHBLJ022716.6+020158            &  18.2 & 0.650  \\
SHBLJ013632.5+390559            & 15.4  & 0.193 &&& SHBLJ023536.6$-$293843          &  17.6 & 0.520 \\
SHBLJ032350.7+071736$^{(a)}$    & 20.3  & /     &&& SHBLJ030416.3$-$283217$^{(c)}$  &  19.3 & /       \\
SHBLJ041112.2$-$394143$^{(a)}$  & 18.8  & /     &&& SHBLJ031422.9+061956            &  18.0 & 0.623    \\
SHBLJ050335.3-111507            & 17.9  & 0.596 &&& SHBLJ035856.1$-$305448          &  18.5 & 0.650   \\
SHBLJ092401.1+053345            & 18.4  & 0.650 &&& SHBLJ042900.1$-$323640          &  17.6 & 0.520   \\
SHBLJ095805.9$-$031740$^{(a)}$  & 19.6  & /     &&& SHBLJ044018.5$-$245933$^{(d)}$  &  19.5 & /         \\
SHBLJ102243.8$-$011302          & 15.5  & 0.202 &&& SHBLJ112348.9+722958            &  18.5 & 0.650   \\
SHBLJ124149.3$-$145558          & 17.3  & 0.455 &&& SHBLJ230634.9$-$110348$^{(e)}$  &  19.2 & /       \\
SHBLJ125015.5+315559$^{(a)}$    & 19.5  & /     &&& SHBLJ235730.0$-$171805          &  17.3 & 0.455    \\
SHBLJ140630.1$-$393509$^{(a)}$  & 19.8  & /     &&&   \\
SHBLJ140630.2+123620            & 20.6  & 0.874 &&&   \\
SHBLJ140659.2+164207            & 18.0  & 0.623 &&&    \\
SHBLJ144506.3$-$032612          & 17.4  & 0.476 &&&    \\
SHBLJ150340.6$-$154113          & 17.5  & 0.497 &&&    \\
SHBLJ151618.6$-$152343          & 18.7  & 0.650 &&&    \\
SHBLJ161204.6$-$043815$^{(a)}$  & 18.9  & /     &&&    \\
SHBLJ175615.9+552217            & 17.6  & 0.520 &&&     \\
SHBLJ205242.7+081038$^{(a)}$    & 19.6  & /     &&&    \\
SHBLJ213151.3$-$251558          & 17.3  & 0.455 &&&     \\
SHBLJ224910.7$-$130002$^{(a)}$  & 18.9  & /     &&&    \\
SHBLJ225147.3$-$320614$^{(a)}$  & 19.0  & /     &&&    \\
SHBLJ230436.8+370507            & 17.8  & 0.570 &&&  \\
SHBLJ230722.0$-$120518          & 18.6  & 0.650 &&&  \\
SHBLJ231028.0$-$371909          & 17.8  & 0.570 &&&    \\
\hline

\end{tabular}
\vspace{0.3cm}
\\
(a): low quality spectrum; (b): no published spectrum, \citet{Nass96}; (c): no published spectrum, \citet{Lon02}; (d): no spectrum, \citet{Schwope00}; (e): no spectrum, \citet{Bauer00}.\\

\end{table*}

A lower limit on the redshift of featureless BL Lacs can then be
determined from the estimate of their minimum jet/galaxy ratio using
the fact that ellipticals have a rather constant luminosity. The
jet/galaxy ratio constrains the apparent magnitude of the host galaxy
from the observed total magnitude of the source, which in turn
constrains the redshift. For our sample we have assumed jet/galaxy
ratios of $1$ and $10$ and have used the relation $V_{\rm gal} = 5.10
\cdot \log{z} + 21.65$ from \citet{Bro93} to estimate redshifts. If
the resulting redshift for a jet/galaxy ratio $=10$ was higher than
$z=0.65$, we concluded that the source was at high redshifts and that a
reasonable lower limit on the redshift could possibly be derived by instead using 
a jet/galaxy ratio $=1$. This new redshift limit, however,
obviously had to be $\ge0.65$. If this was not the case, we concluded
that the Ca H\&K break was at observed wavelengths $\ga6500 \AA$ and chose
a conservative lower limit of $z=0.65$. In practise this means that
for sources with total apparent magnitudes $V\la18.1$, we derived
redshift lower limits assuming a jet/galaxy ratio$=10$, for sources
with $18.1\la V \la 19.8$ we chose $z=0.65$, and for fainter sources we
derived redshift lower limits assuming a jet/galaxy ratio$=1$. We 
applied this method only to sources with high S/N ($\ga20$), high-resolution spectra (26/39 objects), since only these can be reliably
classified as definitely featureless.
The stardard deviation on the relation of \citet{Bro93} is 0.88 in $V$ (see their Fig. 2). 
This translates into an error of  $\sim0.10$ in z.
In Table \ref{featureless} we list the featureless objects of our
survey (39).  The columns are as follows: (1) Sedentary source name; (2)
 the visual apparent magnitude estimated from the APM for the northern hemisphere and from
COSMOS $B_J$ magnitudes, as explained in Paper I; (3) the lower limit
redshift computed with the method explained before.

As can be seen in Fig. \ref{highz}, the spectral optical slope hardens with increasing jet/galaxy ratio. Unfortunately, the measured slope itself cannot be used to improve on the lower limit on z. Since we do not know z, we do not know the rest-frame wavelength and, as can be seen from Fig. 2, the amount of spectral hardening with increasing jet/galaxy ratio differs along the spectrum.

\subsection{Classification}

\label{sedclass}

The criteria for the classification of a radio-loud AGN as a BL Lac
have been continuously revised since the first definition of this
object class by \citet{Str72} and most
recently by some of us \citep{L04}. The
separation of BL Lacs from other types of radio-loud AGN, namely radio
galaxies and quasars, is based on two features in their optical
spectra: the strength of their emission lines and the Ca H\&K
break absorption value. The first is used to distinguish two
intrinsically different classes of radio-loud AGN (BL Lacs and flat spectrum 
radio quasars), the second is used to separate strongly and weakly beamed sources.
Our criteria for classifying BL Lac objects in this survey, as already mentioned in Paper II, 
was defined by \citet{Marcha96} and confirmed later by \citet{L02}.

We have made no effort to separate the weakly from the strongly beamed
sources based on the value of the Ca H\&K break (as suggested by March\~a
et al. 1996 and Landt et al. 2002), since we did not want to bias our
sample against low-luminosity BL Lacs. However, we discuss the
Ca H\&K-break value distribution of our sources in more detail in Sect.
4.2.

We have also investigated how the classification method of \citet{L04} applies to our sample.
We considered all the 169 sources (150 BL Lac and 19 emission line AGN) in the ``HBL zone'' (see Paper II) and used the \OIII~$\lambda 5007$--~\OII~$\lambda 3727$ rest-frame equivalent width plane (see Fig. 4 of \citealt{L04}) to separate our sources into weak- and strong-lined AGN. 

\citet{L04} present evidence of a bimodal \OIII~distribution
in radio-loud AGN and define the two classes as sources with {\it
intrinsically} weak- and strong \OIII~emission lines, respectively. Of
the 19 emission-line AGN in our sample (see Paper II; Table 3), 8
sources have been observed by us, and published equivalent width values
for \OII~and \OIII~are available for a additional 2 sources, namely 
1RXSJ000729.3$+$02405 \citep{For01} and 1RXSJ122044.5$+$69053 \citep{Puch92}. We
classify only two emission-line AGN, 1RXSJ122044.5$+$69053 and
1RXSJ222944.5$-$27553, as weak-lined AGN. Although no information on emission line equivalent widths could
be obtained for the 9/11 sources already known, we are confident that these belong to the class of
strong-lined AGN. All of these sources are classified in \citet{Bauer00} 
as either Seyfert 1 or Seyfert 1.5, which means that they
have strong broad emission lines, a characteristic atypical of
weak-lined AGN \citep{L04}. For sources observed by us without
emission lines, we derived $2\sigma$ non-detection upper limits as
described in Section 3.3., and based on these, we can classify all of these
objects as weak-lined AGN (assuming the same upper limit for both
\OII~and \OIII). We assume that this classification also holds for the 
sources from the literature without emission lines, since those
spectra should have had a similar quality to ours. In summary, following the classification method of \citet{L04}, our
sample contains 152 weak-lined AGN (which we refer to as BL Lacs) and
17 strong-lined AGN.


\subsection{Notes on individual spectra}
\label{notes}

{\bf SHBL J001527.9$-$353639} We classify this object as BL Lac, but its redshift cannot be determined because of the absence of emission and absorption features.

\noindent {\bf SHBL J003334.2$-$192133} This object was already identified as BL Lac by \citet{Bauer00}. We confirm its identification and found a tentative redshift of z=0.610.

\noindent {\bf SHBL J004208.0$+$364112} This object has already been identified as BL Lac by \citet{Bauer00}. We confirm the identification, but its redshift cannot be determined because of the absence of emission and absorption features. 

\noindent {\bf SHBL J012657.2$+$330730} We classify this object as BL Lac, but its redshift cannot be determined because of the absence of emission and absorption features.

\noindent {\bf SHBL J013632.5$+$390559} This object has already been identified as BL Lac by \citet{wei99}. We confirm the identification, but its redshift cannot be determined because of the absence of emission and absorption features.

\noindent {\bf SHBL J032350.7$-$071737} We classify this object as BL Lac, but its redshift cannot be determined because of the absence of emission and absorption features.

\noindent {\bf SHBL J041112.2$-$394143} We classify this object as BL Lac, but its redshift cannot be determined because of the absence of emission and absorption features.

\noindent {\bf SHBL J050335.3$-$111507} We classify this object as BL Lac, but its redshift cannot be determined because of the absence of emission and absorption features.  

\noindent {\bf SHBL J075124.9$+$173051} This object was identified as Seyfert 2 by \citet{wei99} with redshift z=0.185. 
We confirm its redshift but classify it as BL Lac, as its optical spectrum does not show emission lines.
The Ca H\&K break for this object is located in noise, so its measurement is impossible.

\noindent {\bf SHBL J092401.1$+$053345} This object has already been identified as BL Lac candidate by  \citet{Bauer00}. We confirm the identification of BL Lac, but its redshift cannot be determined because of the absence of emission and absorption features.

\noindent {\bf SHBL J095224.0$+$750213} This object was identified as an early type galaxy by \citet{Bauer00} with redshift z=0.181. We classify this object as a BL Lac with a redshift of z=0.179. The Ca H\&K break for this object is located in noise so its measurement is impossible. 

\noindent {\bf SHBL J095805.9$-$031740} We classify this object as BL Lac, but its redshift cannot be determined because of the absence of emission and absorption features.

\noindent {\bf SHBL J102243.8$-$011302} This object has already been identified as BL Lac by \citet{Bauer00}. We confirm the identification, but its redshift cannot be determined because of the absence of emission and absorption features.

\noindent {\bf SHBL J114535.1$-$034001} It was classified as cluster of galaxies (z=0.167) by \citet{Schwope00}; since its X-ray emission is not extended we observed it in order to investigate if this object could be a BL Lac in a cluster. From our spectrum we classify it as a BL Lac and confirm the redshift published by \citet{Schwope00}.

\noindent {\bf SHBL J124149.3$-$145558} This object has already been identified as BL Lac by \citet{PG95}. We confirm the identification, but its redshift cannot be determined because of the absence of emission and absorption features.

\noindent {\bf SHBL J125015.5$+$315559} We classify this object as BL Lac. It was observed under non-photometric conditions and is characterized by a low S/N (see Table \ref{featureless}).

\noindent {\bf SHBL J140630.1$-$393509} We classify this object as BL Lac, but its redshift cannot be determined because of the absence of emission and absorption features.

\noindent {\bf SHBL J140630.2$+$123620} We classify this object as BL Lac, but its redshift cannot be determined because of the absence of emission and absorption features.

\noindent {\bf SHBL J140659.2$+$164207} This object has already been identified as BL Lac by \citet{Bauer00}. We confirm the identification, but its redshift cannot be determined because of the absence of emission and absorption features.

\noindent {\bf SHBL J143917.4$+$393243} This object has already been identified as BL Lac by \citet{Bauer00}. We confirm its identification and we also found its redshift (z= 0.344).

\noindent {\bf SHBL J144506.3$-$032612} This object has already been identified as BL Lac by \citet{Bauer00}. We confirm the identification, but its redshift cannot be determined because of the absence of emission and absorption features.

\noindent {\bf SHBL J150340.6$-$154113} This object has already been identified as BL Lac by \citet{Bauer00}. We confirm the identification, but its redshift cannot be determined because of the absence of emission and absorption features.

\noindent {\bf SHBL J151041.0$+$333504} We classify this object as a BL Lac with redshift z=0.112. After our observation in 1999, \citet{Schwope00} classified this object as BL Lac with z=0.113, thus confirming our identification.

\noindent {\bf SHBL J151618.6$-$152343} This object has already been identified as BL Lac by \citet{Bauer00}. We confirm the identification, but its redshift cannot be determined because of the absence of emission and absorption features.

\noindent {\bf SHBL J153311.3$+$185428} This object has already been identified as BL Lac by \citet{Bauer00}. We confirm its identification and we found also its redshift (z= 0.305).

\noindent {\bf SHBL J161204.6$-$043815} We classify this object as BL Lac, but its redshift cannot be determined because of the absence of emission and absorption features.

\noindent {\bf SHBL J175615.9$+$552217} This object has already been identified as BL Lac by \citet{Bauer00}. We confirm the identification, but its redshift cannot be determined because of the absence of emission and absorption features.

\noindent {\bf SHBL J205242.7$+$081038} We classify this object as BL Lac, but its redshift cannot be determined because of the absence of emission and absorption features.

\noindent {\bf SHBL J213135.4$-$091523} This object has already been identified as BL Lac by \citet{Bauer00}. We confirm its identification and found a tentative redshift of z= 0.449.

\noindent {\bf SHBL J213151.3$-$251558} We classify this object as BL Lac, but its redshift cannot be determined because of the absence of emission and absorption features.

\noindent {\bf SHBL J213852.5$-$205348} This object was classified as candidate BL Lac by \citet{Bauer00}; we confirm its identification and found its redshift (z= 0.290). 

\noindent {\bf SHBL J224910.7$-$130002} This object has already been identified as BL Lac by \citet{Bauer00}. We confirm the identification, but its redshift cannot be determined because of the absence of emission and absorption features.

\noindent {\bf SHBL J225147.3$-$320614} This object has already been identified as BL Lac by \citet{Bauer00}. We confirm the identification, but its redshift cannot be determined because of the absence of emission and absorption features.

\noindent {\bf SHBL J230436.8$+$370507} This object has already been identified as BL Lac by \citet{Cao99}. We confirm the identification, but its redshift cannot be determined because of the absence of emission and absorption features.

\noindent {\bf SHBL J230722.0$-$120518} We classify this object as BL Lac, but its redshift cannot be determined because of the absence of emission and absorption features.

\noindent {\bf SHBL J231028.0$-$371909} We classify this object as BL Lac, but its redshift cannot be determined because of the absence of emission and absorption features.

\noindent {\bf SHBL J235023.2$-$243603} It was classified as cluster of galaxies (z=0.193) by \citet{Col98}; since its X-ray emission is not extended, we observed it in order to investigate if this object could be BL Lacs in cluster.  From our spectrum we classify it as a BL Lac and confirm the redshift published by \citet{Col98}.

\section{Sample properties}
\label{properties}
The full Sedentary HBL sample includes 150 objects and is 100\% spectroscopically identified. 
Redshift have been obtained for 111 objects (74\% of the total sample).

\subsection{Redshift distribution}
\label{redshift}

The sedentary survey redshift distribution has been de-convolved with the appropriate sky coverage. Each bin represents $\sum 1/{\rm Area}(f_{\rm x})$ for all the sources in that bin, where Area$(f_{\rm x})$ is the area accessible at its X-ray flux, divided by the total surface density of sources \citep{L01}.

\begin{figure}[h]
\centering
\includegraphics[height=8cm, width=8cm]{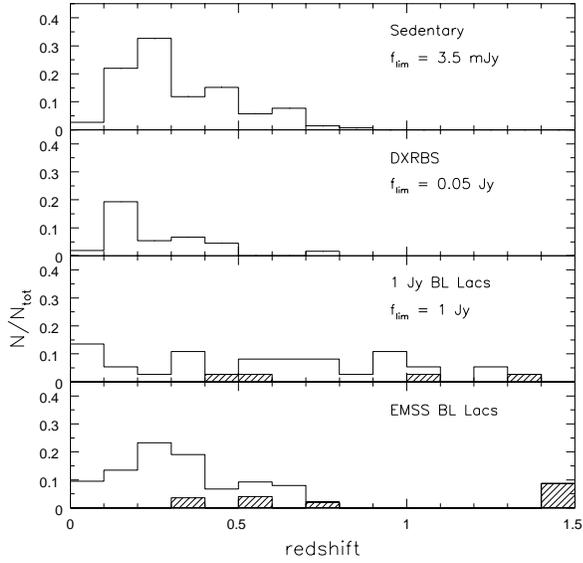}
\caption{Fractional redshift distribution for the 111 sedentary, 17 DXRBS, 32 1 Jy, and 38 
EMSS BL Lacs. The sedentary, DXRBS, and EMSS distributions have been de-convolved with 
the appropriate sky coverages. The hatched areas represent lower limits (1 Jy) and uncertain values (EMSS). 
See text for details.}
\label{dist_z}
\end{figure}

We compared the sedentary survey fractional redshift distribution (see Fig. \ref{dist_z}) with various distributions from other 
BL Lac surveys, namely the complete DXRBS BL Lac sample \citep{Pad07}, the 1 Jy \citep{Sti91, Sti94c, Sto97}, and the EMSS \citep{Rec00} samples. The DXRBS and EMSS redshift distributions have been de-convolved with the appropriate sky coverage.  
Five EMSS redshifts are uncertain, while four 1 Jy redshifts are lower limits (Fig. \ref{dist_z}). Five additional 1 Jy sources have a 0.2 lower limit on their redshift based on non-detection of their host galaxies on the optical images \citep{Sti91}. 
Note that the fraction of BL Lacs with redshift information ranges from $93\%$ and $86\%$ for the EMSS and 1 Jy samples, respectively, to $74\%$ and $71\%$ for the sedentary and the DXRBS, respectively. 

The mean redshift for the four BL Lac samples is  $\langle z \rangle = 0.32$ for the sedentary, 
$\langle z \rangle = 0.26$ for the DXRBS, $\langle z \rangle = 0.46$ for the EMSS, and 
$\langle z \rangle = 0.63$ (including lower limits) for the 1 Jy. The sedentary, DXRBS and EMSS 
samples are peaked at $z=0.3$, $z=0.2$ and $z=0.3-0.4$, respectively, and neither sample includes 
a significant number of $z>0.8$ objects ($\sim 1\%$ in the sedentary, $\sim 9\%$ in the EMSS, and 
$\sim 23\%$ in the DXRBS). By comparison, the 1 Jy BL Lacs have a somewhat surprising flat redshift
distribution out to nearly $z=1.5$, with 10/32 1 Jy BL Lacs at $z >0.8$ and 5 at $z>1$

\subsection{The Ca H\&K break distribution}

\begin{figure}[h]
\centering
\includegraphics[height=8cm, width=8cm]{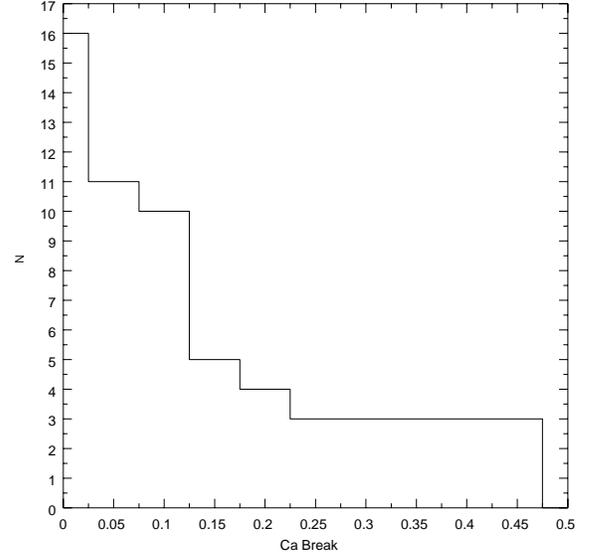}
\caption{Ca H\&K break distribution of 61 sources from sedentary survey (from literature and from our spectroscopic campaign)}
\label{dist_Ca}
\end{figure}

Figure \ref{dist_Ca} shows the Ca H\&K break distribution of those HBL (61) from the sedentary survey for which we could take measurements from the literature or from our spectroscopic identifications.

\citet{L02} showed that the Ca H\&K break value of low-luminosity, 
radio-loud AGN is a suitable statistical orientation indicator and can
be used to roughly separate the strongly from the weakly beamed sources. This
feature  is on average $\sim 0.5$ in normal
non-active ellipticals \citep{Dre87} and is decreased by the beamed
non-thermal jet emission in blazars. 
We measured the Ca H\&K break value for our sources (in spectra plotted as
$f_{\lambda}$ versus $\lambda$) and list these in Table \ref{prop}.
The error represents the 1 $\sigma$ limit and was computed based on the
S/N blue ward and red ward of the feature.

Most of sedentary objects have Ca H\&K break = 0 or below 0.25. This property was expected, since our sample is constituted by a particular class of BL Lacs, the HBLs, characterized by a synchrotron emission peak located at high energies (UV/X-ray energy band) for which the dilution of host galaxies optical light is very high.

\subsection{Optical slopes Vs. radio luminosities}

\begin{figure}[h]
\centering
\includegraphics[height=8cm, width=8cm]{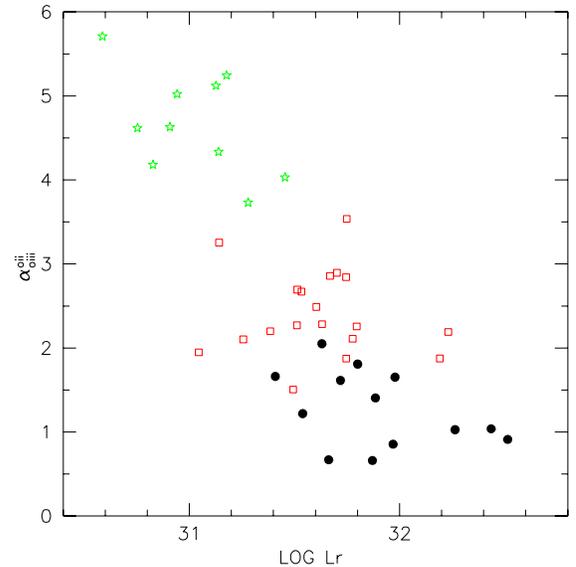}
\caption{Optical spectral slopes ($\alpha^{oii}_{oiii}$) for 38 sedentary objects versus their 1.4 GHz radio luminosity ($L_r$). 
         Different symbols indicate Ca H\&K break values in the ranges $C\le 0.05$ (filled circles), $0.05<C<0.25$ (open 
         squares), and $C \ge 0.25$ (stars). 
        }
\label{alpha_Lr}
\end{figure}

In Fig. \ref{alpha_Lr} we have plotted the optical spectral slopes between the rest frame frequencies of \OII and \OIII ($\alpha^{oii}_{oiii}$) of 41 sedentary objects (38 sources with spectra observed by us, 2 sources with spectra from the \citet{Sloan01} and 1 object from the DXRBS, \citealt{L01}) versus their radio luminosity ($L_r$) at 1.4GHz using different symbols for Ca H\&K break values in the ranges $C\le 0.05$, $0.05<C<0.25$, and $C \ge 0.25$. 
The remaining 3 objects with redshift observed by us were excluded from this sample because their Ca H\&K break is located in noisy parts of the spectra making its measurement impossible.

\citet{L02} showed that the Ca H\&K break value decreases with increasing jet powers, concluding that the Ca H\&K break value of BL Lacs and radio galaxies is a suitable indicator of orientation.
We indeed find that this also applies to the sedentary survey. As Fig. \ref{alpha_Lr} shows, there is a correlation between $\alpha^{oii}_{oiii}$ and $L_r$, reflecting the fact that for more intrinsically powerful and/or beamed sources (in the radio band), i.e. objects with stronger and/or more beamed non thermal emission, the optical light is dominated by the jet and is therefore characterized by a harder spectrum.

\section{Summary and conclusions}
\label{conc}

We have presented the results of a dedicated optical spectroscopic campaign of the multi-frequency sedentary survey, 
a flux-limited and statistically well-defined sample of 150 high-energy peaked BL Lacertae objects.
Our program, carried out with the ESO 3.6m, the KPNO 4m, and the TNG optical telescopes, led to the spectroscopic 
identification of {\it all} sources in the sample.

In this paper we have presented optical spectra for 76 sources, 50 of which are new BL Lac objects, 18 are sources 
previously known to be BL Lacs but without redshift determination, and 8 are broad emission-line AGNs. 
We determined 36 redshifts out of the 50 new BL Lacs and 5 new redshifts for the previously known objects. The redshift 
distribution of the complete sample is presented and compared with that of other BL Lacs samples. 
For 26 sources without recognizable absorption features, we calculated lower limits to the redshift using a method 
based on simulated optical spectra with different ratios between jet and galaxy emission.

For a subsample of 38 object with high-quality spectra, we presented the measured Ca H\&K break values, and find a 
correlation between the optical spectral slope, 
the 1.4 GHz radio luminosity, and the Ca H\&K break, 
indicating that for powerful/beamed sources the optical light is dominated by the non-thermal emission from the jet.

The main cosmological properties, such as the luminosity function and the cosmological evolution of the sample, are studied 
in detail in \citet{paperIV}.

\section*{Acknowledgements}
\label{ackn}

This work is partly based on optical spectroscopy observations performed at the
European Southern Observatory, La Silla, Chile, (Proposals ESO n. 67.B-0222(A), 71.B-0582(A),
and 71.B-0582(B)), Telescopio Nazionale Galileo, La Palma, Canary Islands
(proposals AOT5/02A, AOT6/02B, AOT7/03A), and Kitt Peak National Observatory.
We acknowledge ESO, TNG and KPNO personnel for their assistance during the observing runs.
This research has also made use of data taken from the NASA/IPAC Extragalactic Database (NED) and 
the ESO on-line Digitized Sky Survey on-line services.

\bibliography{paper_references.bib}
\appendix
\section{68 sedentary survey spectra}
\input{spectra.tex}

\section{spectra of the 8 AGNs excluded from the sample}
Here we present the optical spectra for the remaining 8 objects observed by us that were broad emission lines AGN 
and because of that they have been excluded from the sample.

\vspace*{1cm}

The wavelength in \AA~is plotted on the x-axis while the
y-axis gives the flux $f_\lambda$ in units of $10^{-17}$ erg
cm$^{-2}$ s$^{-1}$ \AA$^{-1}$.

\begin{figure*}
\includegraphics*[height=26cm, width=18cm]{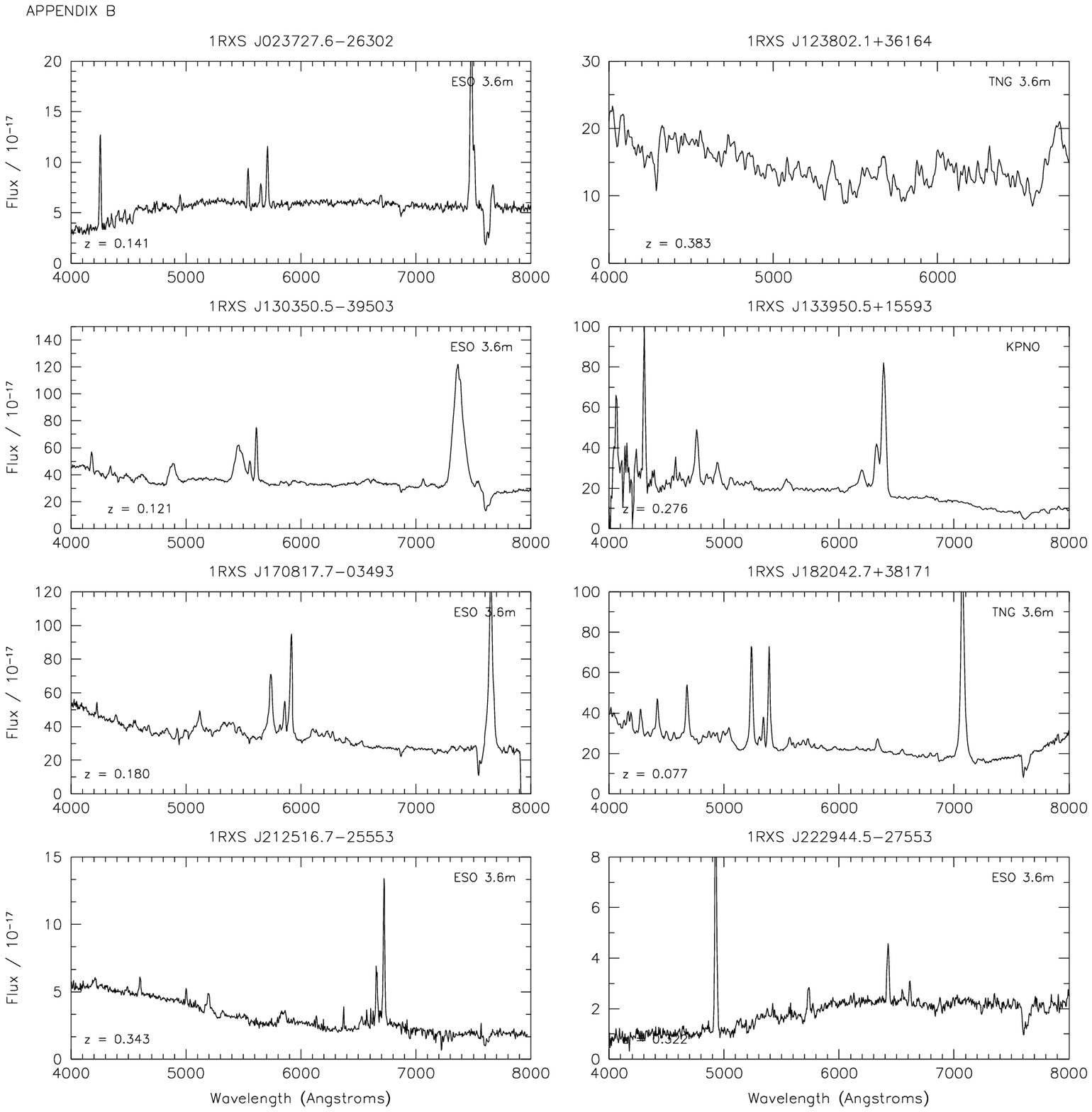}
\end{figure*}

\end{document}

%% file: spectra.tex
\label{appspectra}
We present here the optical spectra for the 68 Sedentary HBL objects observed during the Sedentary identification campaign from 1999 to 2003 and discussed in this paper.

\vspace*{1cm}

The wavelength in \AA~is plotted on the x-axis while the 
y-axis gives the flux $f_\lambda$ in units of $10^{-17}$ erg 
cm$^{-2}$ s$^{-1}$ \AA$^{-1}$.

\begin{figure*}
\includegraphics*[height=26cm, width=18cm]{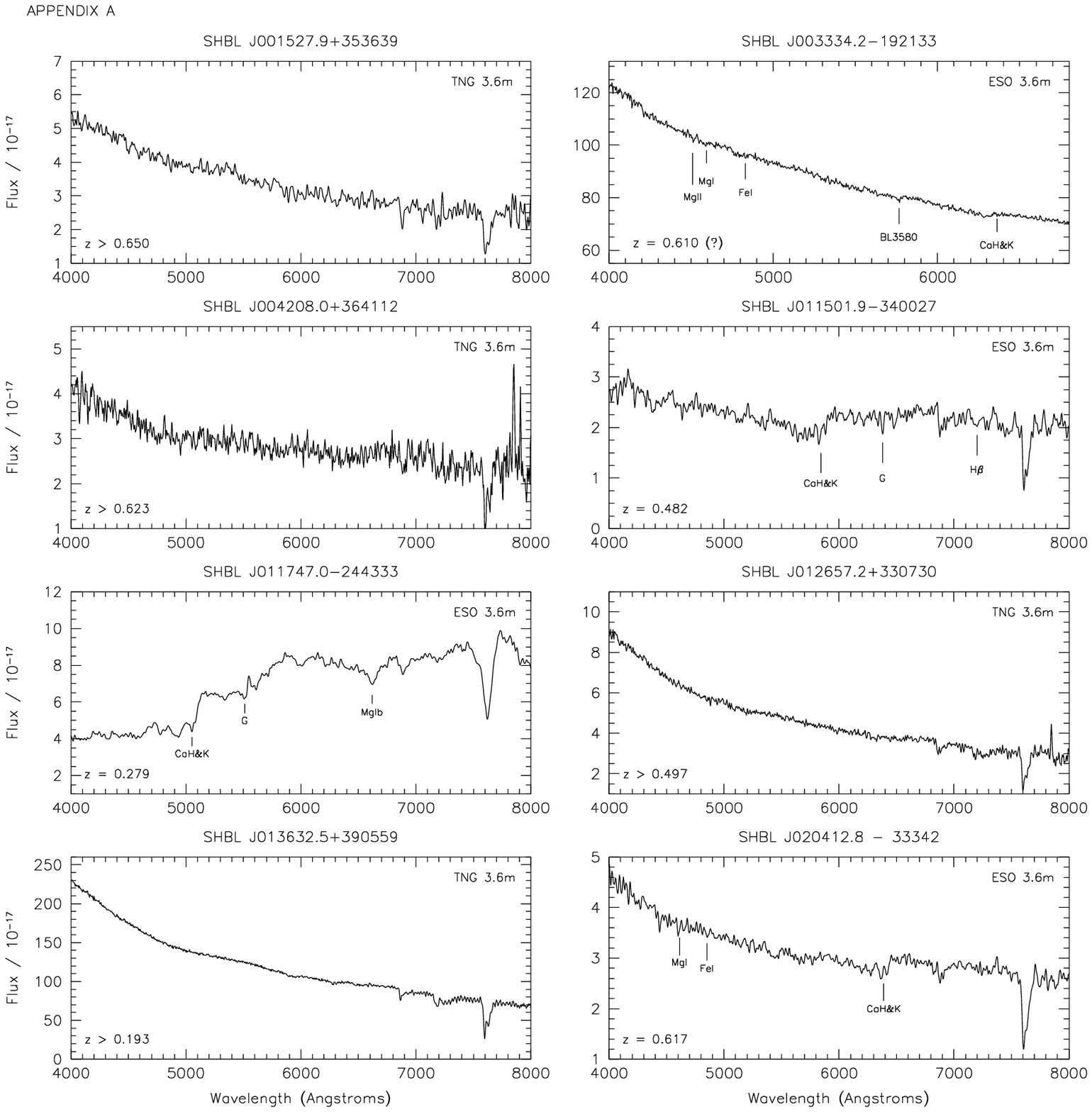}
\end{figure*}

\begin{figure*}
\includegraphics*[height=26cm, width=18cm]{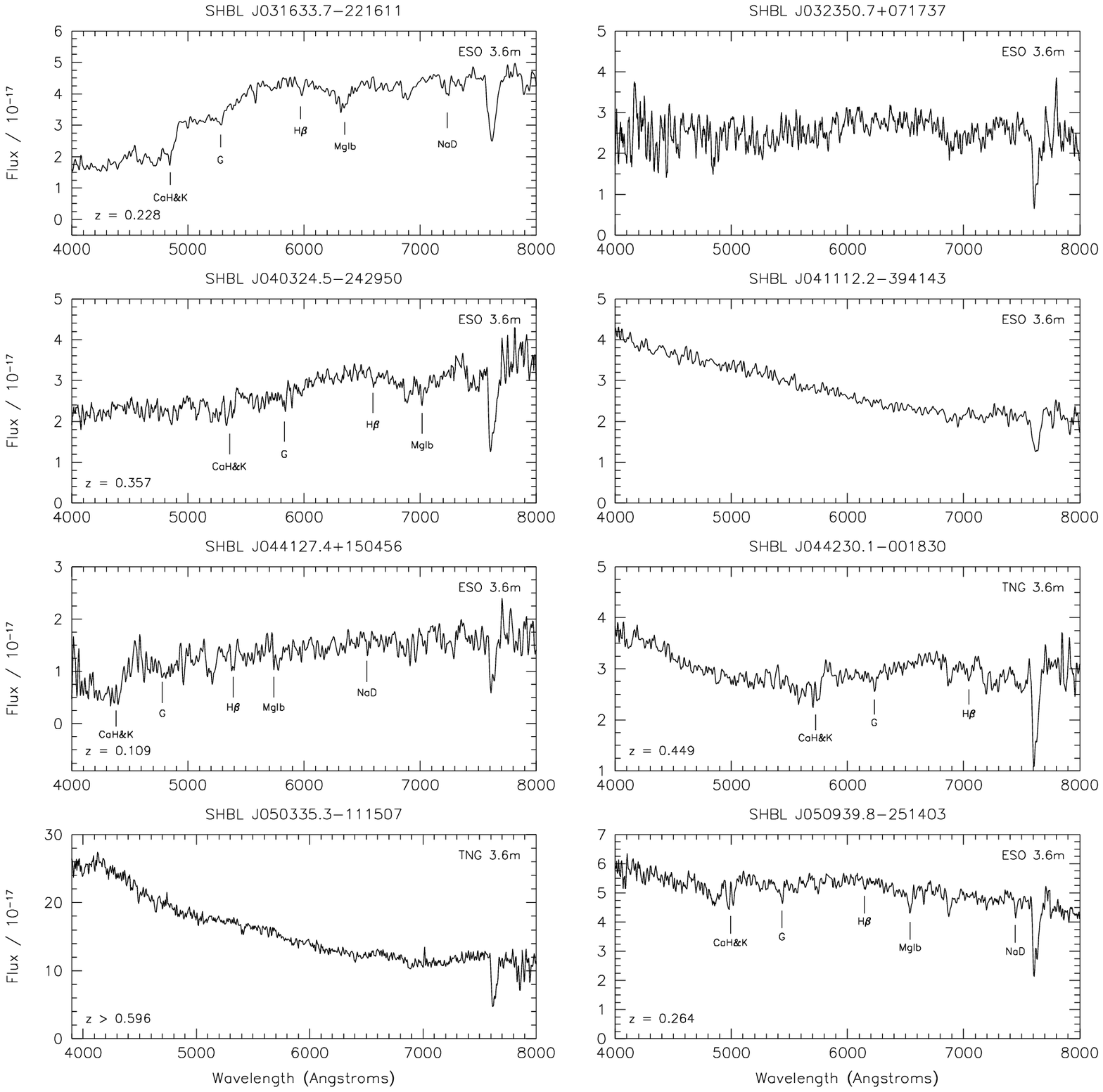}
\end{figure*}

\begin{figure*}
\includegraphics*[height=26cm, width=18cm]{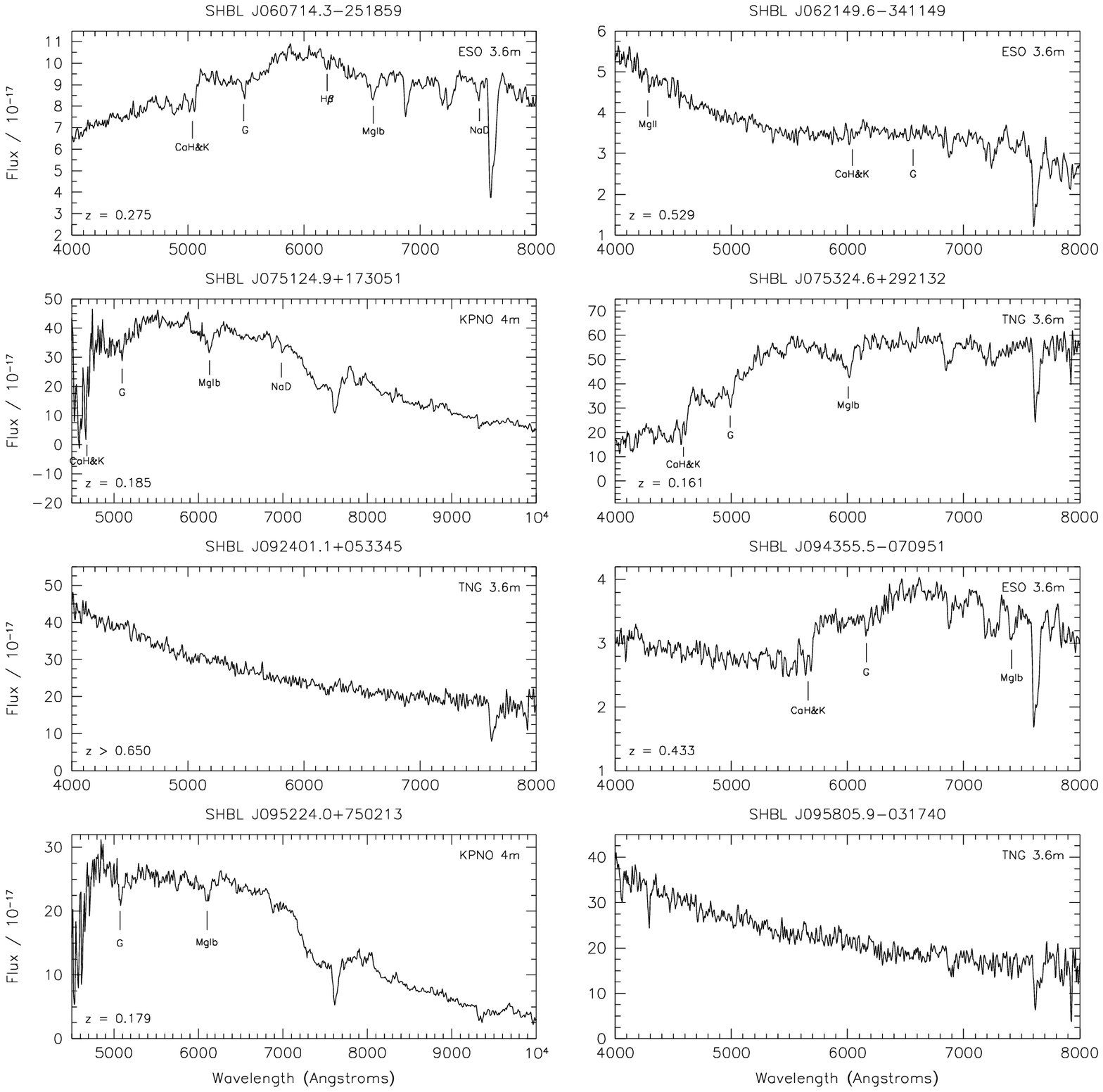}
\end{figure*}

\begin{figure*}
\includegraphics*[height=26cm, width=18cm]{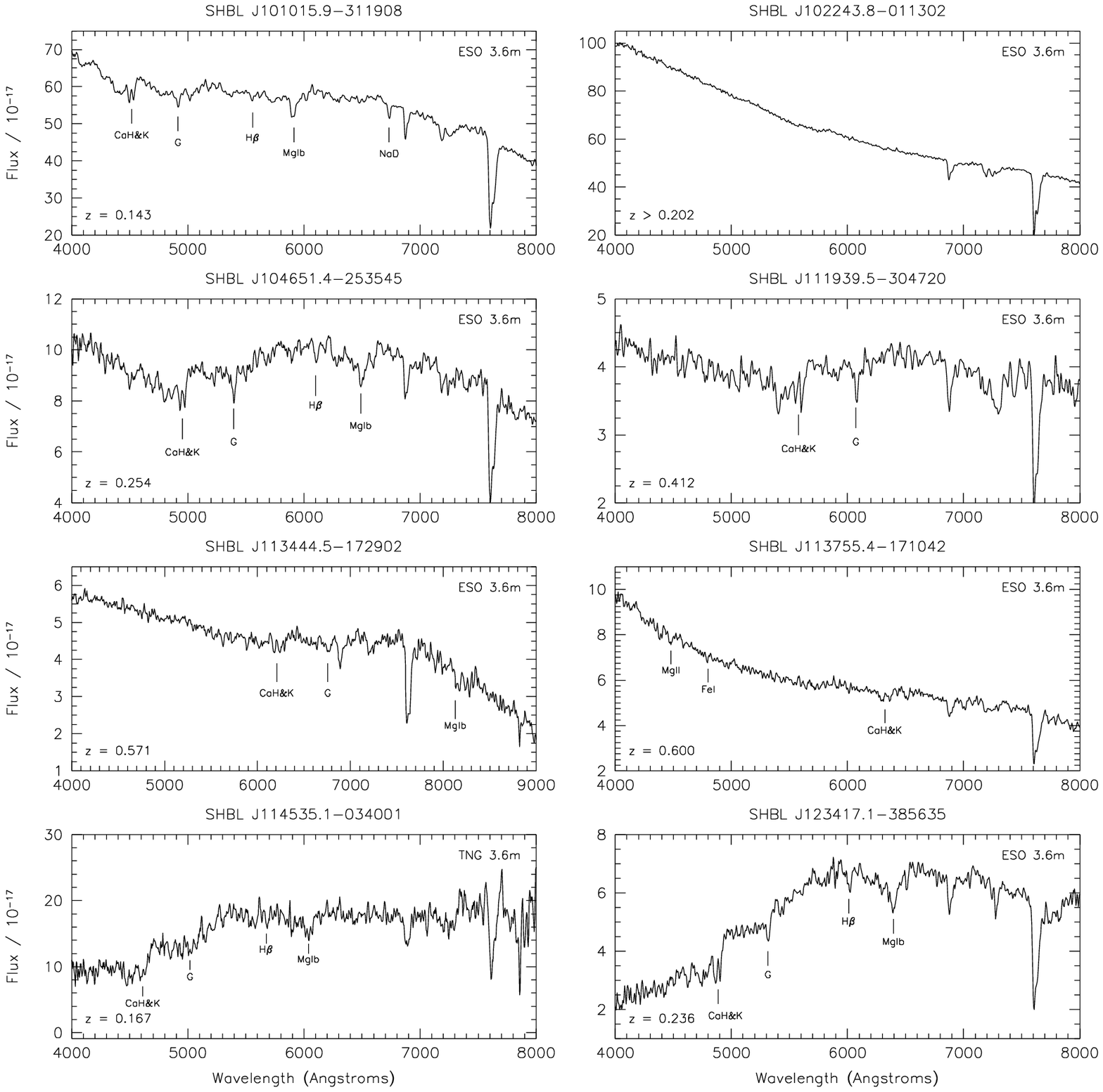}
\end{figure*}

\begin{figure*}
\includegraphics*[height=26cm, width=18cm]{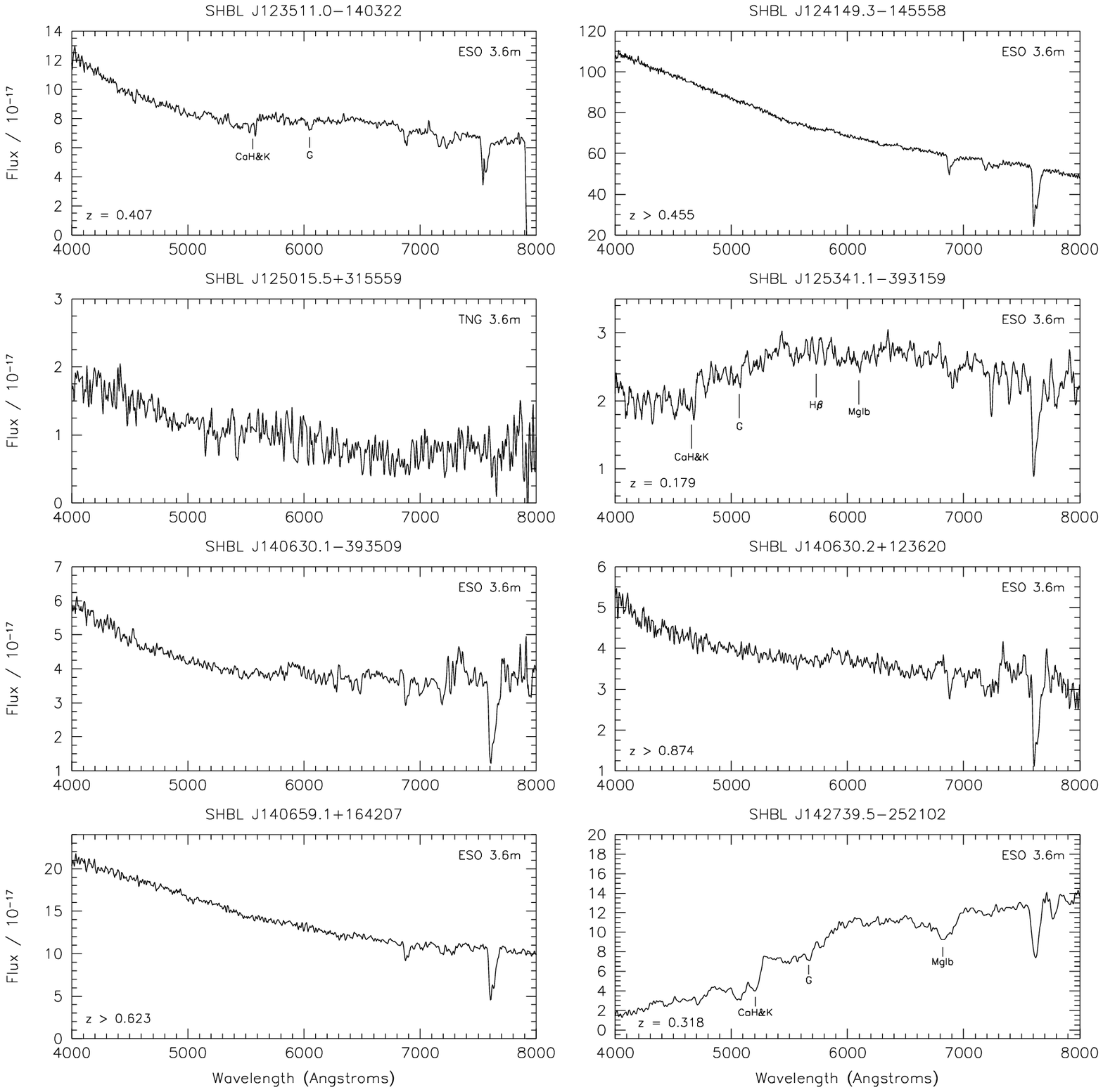}
\end{figure*}

\begin{figure*}
\includegraphics*[height=26cm, width=18cm]{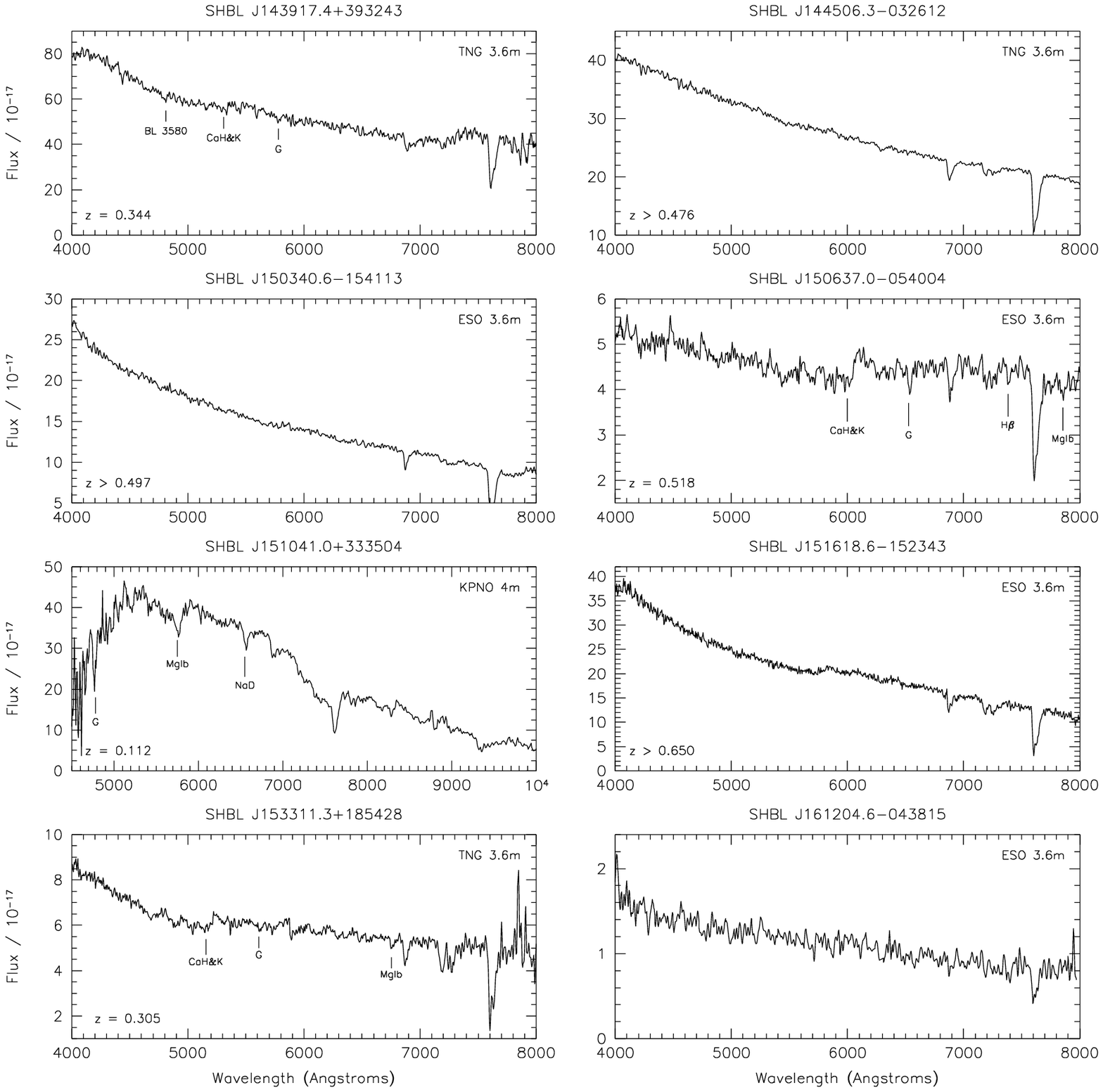}
\end{figure*}

\begin{figure*}
\includegraphics*[height=26cm, width=18cm]{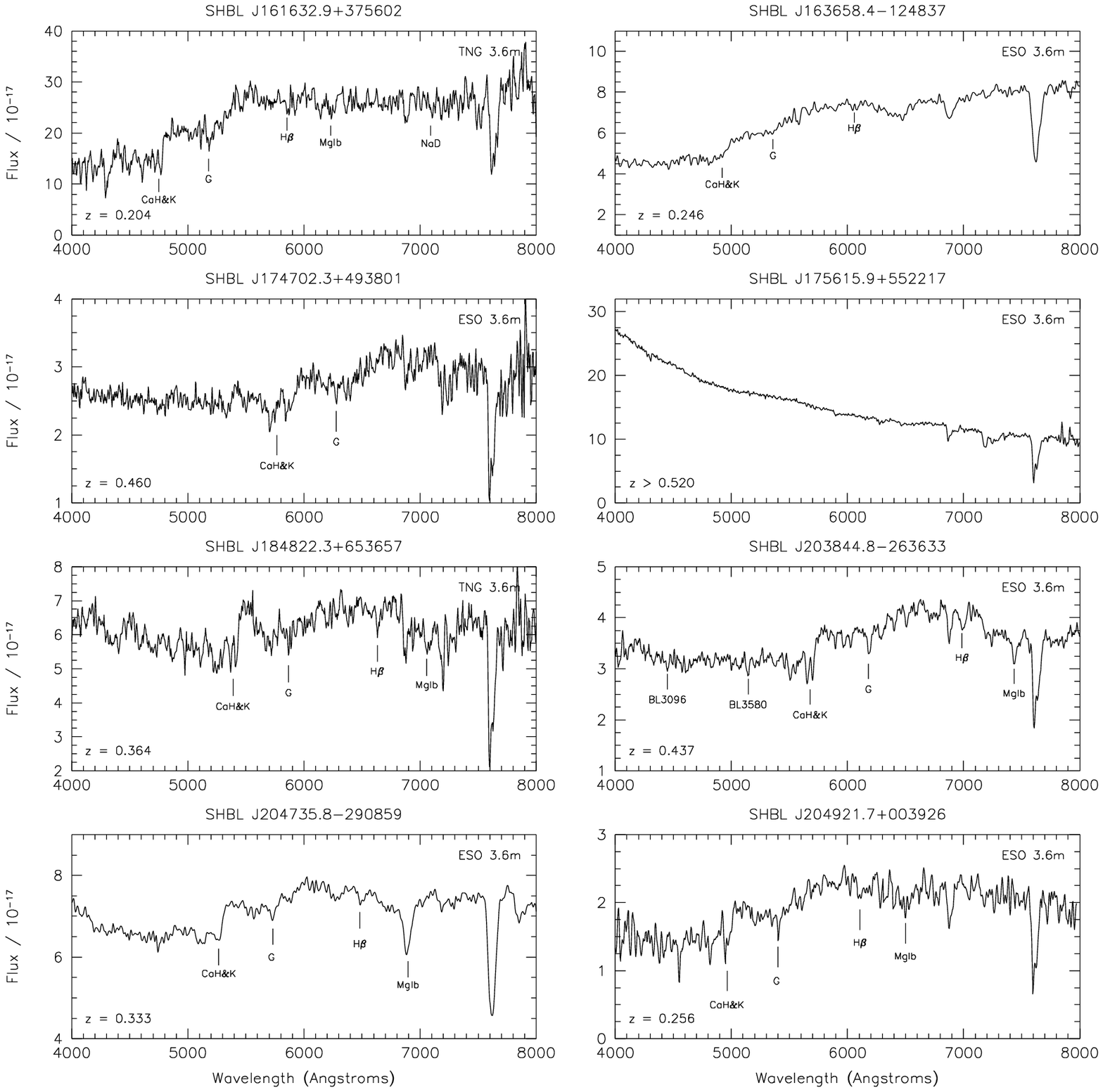}
\end{figure*}

\begin{figure*}
\includegraphics*[height=26cm, width=18cm]{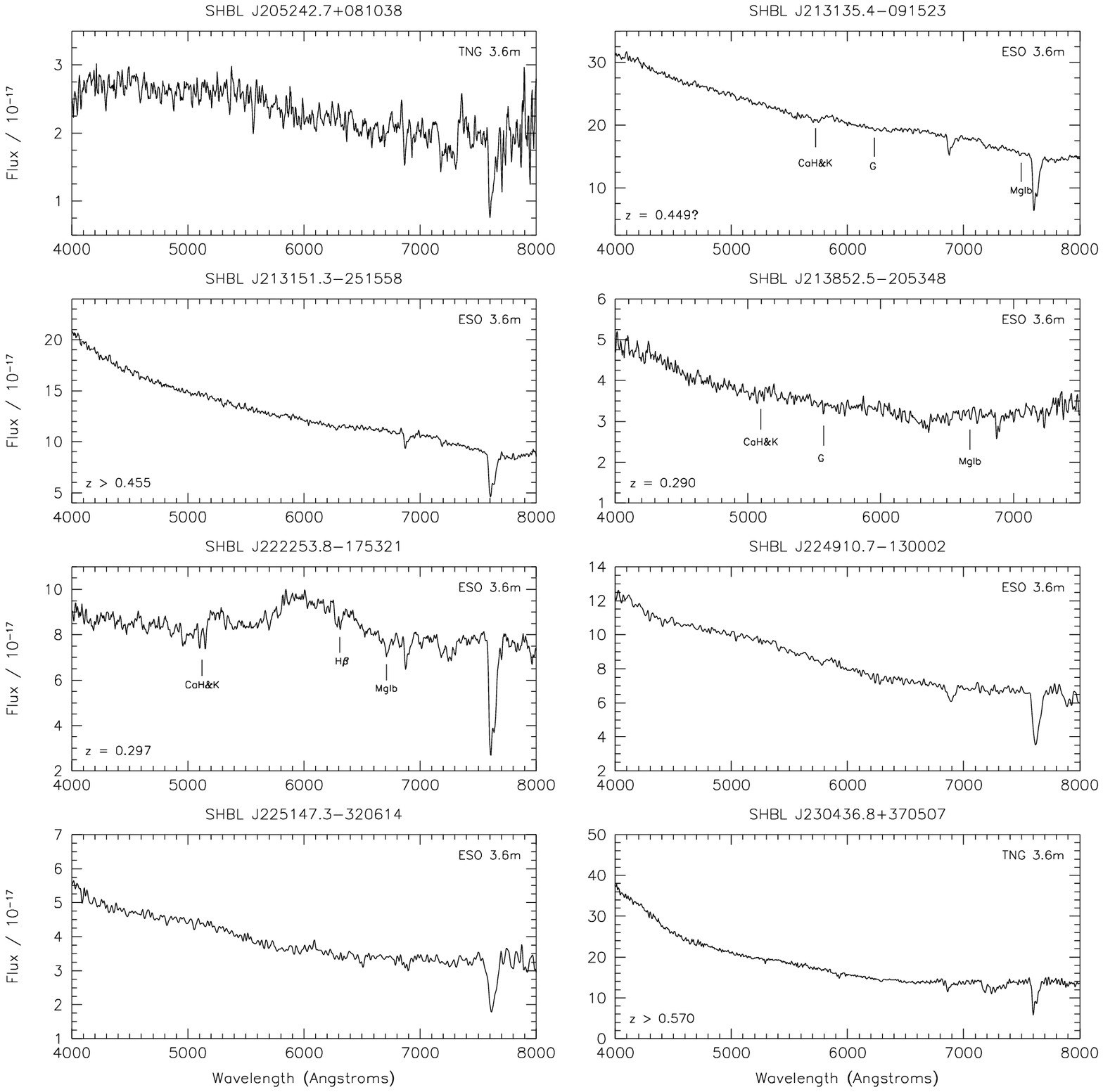}
\end{figure*}

\begin{figure*}
\includegraphics*[height=26cm, width=18cm]{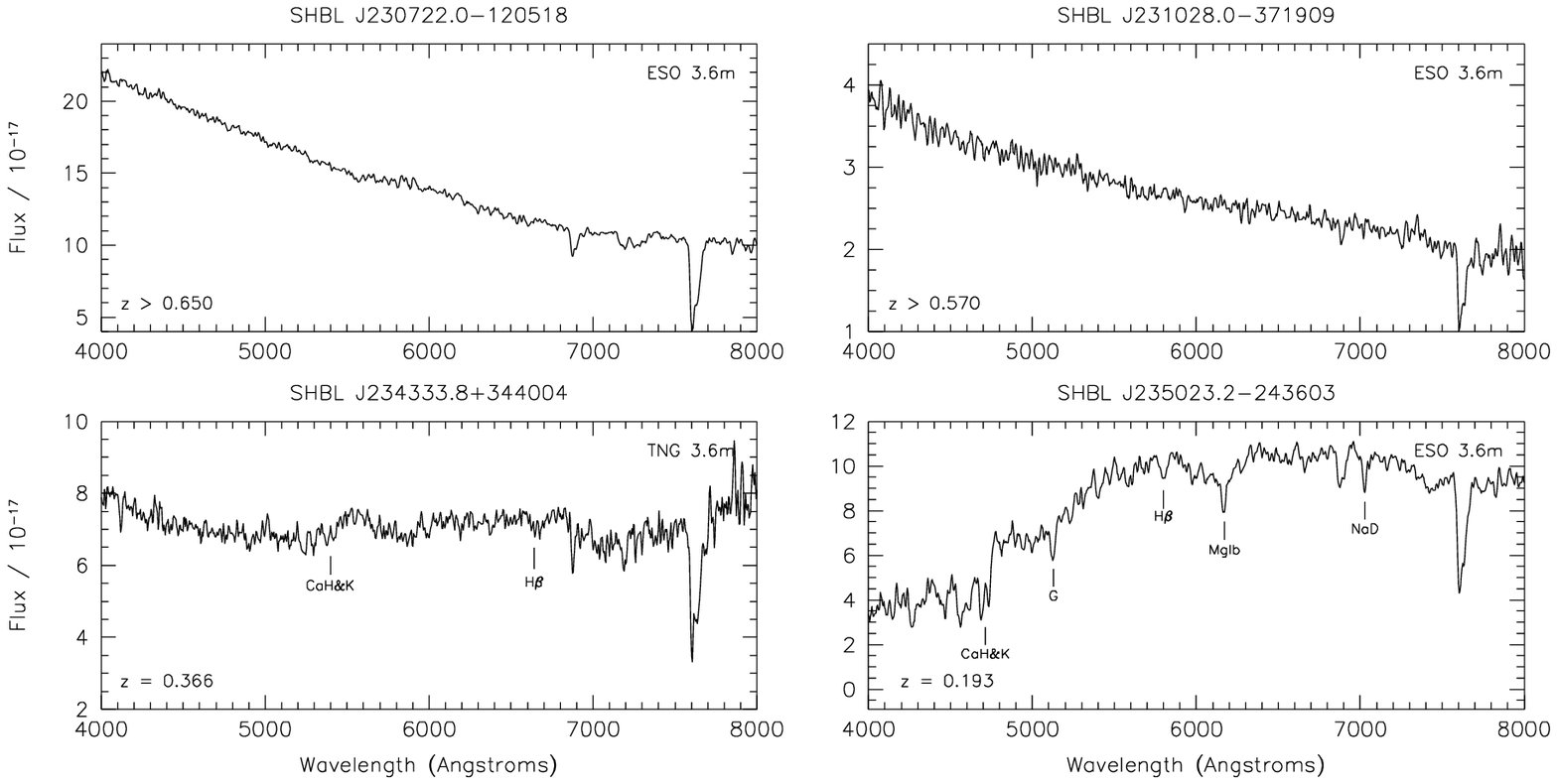}
\end{figure*}